\documentclass[fleqn,usenatbib]{mnras}

\usepackage{newtxtext,newtxmath}

\usepackage[T1]{fontenc}

\DeclareRobustCommand{\VAN}[3]{#2}
\let\VANthebibliography\thebibliography
\def\thebibliography{\DeclareRobustCommand{\VAN}[3]{##3}\VANthebibliography}

\usepackage{hyperref}
\usepackage{natbib}
\usepackage{xspace}
\usepackage{xcolor}
\usepackage{parskip}
\usepackage{CJKutf8}
\newcommand{\CNnames}[1]{{\begin{CJK}{UTF8}{gbsn}~(#1)~\end{CJK}}}

\usepackage{graphicx}	
\usepackage{amsmath}	

\title[Optical spectroscopic signatures of EV]{Optical spectroscopic signatures of the red giant evolutionary state}

\author[E.~Wang]{Ella Xi~Wang$^{1, 2,3}$\thanks{Email: xi.wang@astro.su.se}, 
Melissa~Ness$^{2}$, 
Thomas~Nordlander$^{2,3,4}$, Andrew~R.~Casey$^{5,3,6}$, Sarah~Martell$^{3,7}$, \newauthor Marc~Pinsonneault$^{8}$, Xiaoting~Fu~\CNnames{符晓婷}$^{9}$, Dennis~Stello$^{3,7,10}$, Claudia~Reyes$^{2}$, Marc~Hon$^{11}$, \newauthor Madeleine~McKenzie$^{2,12}$, Mingjie~Jian~\CNnames{简明杰}$^{1}$, Jie~Yu~\CNnames{余杰}$^{2}$,
Sven~Buder$^{2,3}$, Karin~Lind$^{1}$, \newauthor Joss~Bland-Hawthorn$^{3,10}$, Daniel~B.~Zucker$^{3,13,14}$, Pradosh~Barun~Das$^{3,13,14}$, Richard~de~Grijs$^{13,14,15}$ \newauthor and Michael~Hayden$^{16}$
\\
$^1$Department of Astronomy, Stockholm University, AlbaNova University Center, SE--106 91 Stockholm, Sweden \\
$^2$Research School of Astronomy and Astrophysics, Australian National University, Canberra, ACT 2611, Australia \\
$^3$ARC Centre of Excellence for All Sky Astrophysics in 3 Dimensions (ASTRO 3D), Australia \\
$^4$Theoretical Astrophysics, Department of Physics and Astronomy, Uppsala University, Box 516, 751 20 Uppsala, Sweden \\
$^5$School of Physics \& Astronomy, Monash University, Melbourne, VIC 3800, Australia \\
$^6$Center for Computational Astrophysics, Flatiron Institute, 162 Fifth Avenue, New York, NY 10010, USA \\
$^7$School of Physics, University of New South Wales, Sydney, NSW 2052, Australia \\
$^8$Department of Astronomy, The Ohio State University, 140 West 18th Ave, Columbus OH 43210, USA \\
$^9$Purple Mountain Observatory, Chinese Academy of Sciences, Nanjing 210023, PR China \\
$^{10}$Sydney Institute for Astronomy (SIfA), School of Physics, University of Sydney, Sydney, NSW 2006, Australia \\
$^{11}$Kavli Institute for Astrophysics and Space Research, Massachusetts Institute of Technology, Cambridge, MA 02139, USA \\
$^{12}$Carnegie Science Observatories, 813 Santa Barbara St., Pasadena, CA 91101, USA \\
$^{13}$ School of Mathematical and Physical Sciences, Macquarie University, Balaclava Road, Sydney, NSW 2109, Australia \\
$^{14}$ Astrophysics and Space Technologies Research Centre, Macquarie University, Balaclava Road, Sydney, NSW 2109, Australia \\
$^{15}$ International Space Science Institute--Beijing, 1 Nanertiao, Zhongguancun, Hai Dian District, Beijing 100190, China \\
$^{16}$ Homer L. Dodge Department of Physics \& Astronomy, University of Oklahoma, 440 W. Brooks St., Norman, OK 73019, USA\\
}

\date{Accepted XXX. Received YYY; in original form ZZZ}

\pubyear{\the\year{}}

\begin{document}
\label{firstpage}
\pagerange{\pageref{firstpage}--\pageref{lastpage}}
\maketitle

\newcommand{\total}{5949\xspace}
\newcommand{\rgb}{4065\xspace}
\newcommand{\rc}{1884\xspace}
\newcommand{\pairs}{786\xspace}
\newcommand{\rgbpairs}{247\xspace}
\newcommand{\rcpairs}{650\xspace}
\newcommand{\ha}{H$_{\alpha}$\xspace}
\newcommand{\hb}{H$_{\beta}$\xspace}
\newcommand{\teff}{\rm{T}_{\rm{eff}}}
\newcommand{\logg}{\log(g)}
\newcommand{\feh}{[\rm{Fe}/\rm{H}]}
\newcommand{\mgfe}{[\rm{Mg}/\rm{Fe}]}
\newcommand{\snr}{{\rm S}/{\rm N}}
\newcommand{\vmic}{v_{\rm{mic}}}
\newcommand{\vsini}{v\sin(i)}
\newcommand{\mass}{M}
\newcommand{\radius}{R}
\newcommand{\kms}{{\rm km}\,{\rm s}^{-1}}
\newcommand{\cms}{{\rm cm}\,{\rm s}^{-2}}
\newcommand{\dnu}{\Delta \nu}
\newcommand{\numax}{\nu_{\rm{max}}}
\newcommand{\muhz}{\mu{\rm Hz}}
\newcommand{\msun}{\mass_{\odot}}
\newcommand{\galah}{GALAH\xspace} 
\newcommand{\asfgrid}{\textsc{AsfGrid}\xspace} 
\newcommand{\pysme}{\textsc{PySME}\xspace}
\newcommand{\korg}{\textsc{Korg}\xspace}
\newcommand{\moog}{\textsc{MOOG}\xspace}
\newcommand{\pymoogi}{\textsc{pyMOOGi}\xspace}
\newcommand{\atlas}{\textsc{ATLAS9}\xspace}
\newcommand{\pykmod}{\textsc{PyKMOD}\xspace}
\newcommand{\ctwo}{\rm{C}_2}
\newcommand{\cisotope}{^{12}\rm{C}/^{13}\rm{C}}
\newcommand{\cn}{[\rm{C}/\rm{N}]}
\newcommand{\xmark}{}
\newcommand{\edit}[1]{#1}
\newcommand{\mkn}[1]{\textcolor{purple}{MKN: #1}}
\newcommand{\SB}[1]{\textcolor{orange}{SB: #1}}

\newcommand{\xt}[1]{[\textit{\textcolor{orange}{xiaoting: #1}}]}

\begin{abstract}
Modern spectroscopic surveys output large data volumes. Theoretical models provide a means to transform the information encoded in these data to measurements of physical stellar properties. However, in detail the models are incomplete and simplified, and prohibit interpretation of the fine details in spectra. Instead, the available data provide an opportunity to use data-driven, differential analysis techniques, as a means towards understanding spectral signatures. We deploy such an analysis to examine core helium-fusing red clump (RC) and shell hydrogen-fusing red giant branch (RGB) stars, to uncover signatures of evolutionary state imprinted in optical stellar spectra. We exploit \pairs pairs of RC and RGB stars from the GALAH survey, chosen to minimise spectral differences, with evolutionary state classifications from TESS and K2 asteroseismology. We report sub-percent residual, systematic spectral differences between the two classes of stars, and show that these residuals are significant compared to a reference sample of RC$-$RC and RGB$-$RGB pairs selected using the same criteria. First, we report systematic differences in the Swan ($\ctwo$) band and CN bands caused by stellar evolution and a difference in mass, where RGB stars at similar stellar parameters have higher masses than RC stars. Secondly, we observe systematic differences in the line-width of the \ha and \hb lines caused by a difference in microturbulence, as measured by \galah, where we measure higher microturbulence in RC stars than RGB stars. This work demonstrates the ability of large surveys to uncover the subtle spectroscopic signatures of stellar evolution using model-free, data-driven methods. \\
\end{abstract}




\section{Introduction}
Distinguishing between hydrogen-shell-fusing red giant branch (RGB) stars and helium-core-fusing red clump (RC) stars is important to a myriad of science cases. For example: the study of lithium rich giants and their enhancement mechanisms \citep{singh2019, casey2019, martell2021, Yan2021, Zhou2022, tayar2023}, building the local cosmic distance ladder \citep{stanek1998, hawkins2017}, building constrained stellar samples to study the low and high $\alpha$ disk \citep{lu2022}, and mass loss on the RGB \citep{charbonnel2005, howell2024}.

Asteroseismology, the study of internal stellar structure through oscillations, provides an unambiguous way of distinguishing between RC and RGB stars \citep{montalban2010, bedding2011, mosser2011, mosser2012}. Red giants are observed to have solar-like oscillations \citep{frandsen2002, deridder2006, barban2007}. These oscillations are caused by pressure or acoustic (p) modes, which probe the convective envelope; and gravity (g) modes, which probe the core. Whilst g-modes contain information on the stellar core, they usually cannot propagate to the surface of the star and are thus not observable. However, p- and g-modes can couple to form mixed modes \citep{chaplin2013}. Mixed modes carry properties from the stellar core to the surface of the star, allowing us to distinguish between different evolutionary states \citep{beck2011}.

Asteroseismology requires high temporal frequency radial velocity or photometric observations in order to resolve mixed modes \citep{hon2018}. Although great strides have been made for asteroseismic surveys \citep{stello2013, elsworth2017, hon2018}, the number of stars for which we have asteroseismic measurements pales in comparison to the number of stars with spectroscopic data \citep{abdurrouf2022, buder2024}. Using spectroscopy, we are able to measure effective temperature ($\teff$), surface gravity ($\logg$), and metallicity ($\feh$) to high accuracy and precision. However, stellar evolutionary tracks at the same $\feh$ but different mass can overlap in $\teff$-$\logg$ space. Although RC and RGB stars can be separated if the star's mass is known, it is difficult to infer masses for giant stars to a high accuracy and precision using only spectroscopic data.

Machine Learning (ML) algorithms have been employed to expand the number of stars with an evolutionary state classification. Data-driven ML methods that leverage precision asteroseismic measurements have shown success in learning evolutionary state directly from both APOGEE and LAMOST stellar spectra \citep{hawkins2018, ting2018, casey2019, lu2022}. This is achieved through label transfer of asteroseismic parameters to spectra, allowing us to classify the evolutionary state using spectroscopic information. With these ML algorithms achieving 93-98\% accuracy, it is clear that there is an evolutionary state signal in stellar spectra.

Differences in the spectroscopy of RC and RGB stars have been linked to molecular features related to carbon \citep{masseron2017, hawkins2018, banks2023, banks2024}, with the so-called ``deep mixing'' thought to be the cause of this link. \edit{At the end of the first dredge up, the convective envelope recedes to the surface, leaving behind a chemical discontinuity (the so-called $\mu$--barrier) which prevents further mixing of material between the stellar core and surface. During evolution up the red giant branch, the hydrogen-fusing shell expands outwards and contacts this $\mu$--barrier, effectively removing it. When this occurs, the star temporarily drops in luminosity, causing what we observe as the RGB bump. The removal of this chemical discontinuity beyond the RGB bump allows further mixing between the stellar core and surface \citep{sweigart1979, charbonnel1994, charbonnel2007, charbonnel2010, lagarde2012}. }Deep mixing has been observed on the RGB beyond the RGB bump \citep{gilroy1989, charbonnel1998, gratton2000, recioblanco2007, martell2008a, martell2008b, masseron2015, lagarde2019, shetrone2019, roberts2024}: it depletes C, enhances N, and reduces the $\cisotope$ ratio, with the amount depending on the star's mass and metallicity \citep{lagarde2012}. As a result, RC stars of similar mass and metallicity have lower $\cn$ and $\cisotope$ compared to RGB stars, causing an observable difference in carbon-related molecular features in the spectrum. Although existing studies find differences in CN, CH, and CO molecular lines \citep{hawkins2018, banks2023, banks2024}, these studies make use of data-driven models.

There are now a number of large spectroscopic surveys with complementary classifications of evolutionary state from asteroseismic surveys for subsets of stars \citep{pinsonneault2014, pinsonneault2025, buder2024}. Galactic Archaeology with HERMES (\galah) is a spectroscopic survey, with the main mission of studying Galactic archaeology through chemical tagging \citep{desilva2015}. In Data Release 4 \citep{buder2024}, \galah incorporated additional fields that observe asteroseismic targets from the K2 \citep{howell2014} and TESS \citep{ricker2014} surveys.   

In this paper, we study the evolutionary state signal in optical stellar spectra using the GALAH, TESS, and K2 surveys in a data-driven but model-free way. In Section~\ref{sec:data}, we clean and crossmatch the survey data used in this work. We take pairs of RC and RGB stars in order to limit stellar parameter influence, discussed in Section~\ref{sec:method}. In Section~\ref{sec:results}, we show the optical spectroscopic signature of stellar evolutionary state, and link this signal to physical stellar parameters in Section~\ref{sec:discussion}. Lastly, we summarise our findings in Section~\ref{sec:conclusion}.

\section{Data}
\label{sec:data}

\subsection{Spectroscopic catalogue}
In this work, we use measured parameters and abundances as well as observed spectra drawn from the million-star \galah survey. \galah is a stellar spectroscopic survey, using the 2dF fibre positioner \citep{lewis2002} on the 3.9\,m Anglo-Australian Telescope at Siding Spring Observatory. Spectra are taken with the HERMES instrument, which has a resolution of $R=28\,000$ over 4 CCDs, covering wavelengths: $4713$--$4903$\,\AA, $5648$--$5873$\,\AA, $6478$--$6737$\,\AA, and $7585$--$7887$\,\AA\ \citep{barden2010, brzeski2011, heijmans2012, farrell2014, sheinis2015}. Data Release 4 (DR4) measured stellar parameters and abundances for $\sim 900\,000$ unique stars, these parameters along with reduced spectra are published in \citet{buder2024}. 

We remove anomalous stars, flagged stars, and low $\snr$ stars. The \galah catalogue flags stars with issues in the data analysis. In particular, the flags \texttt{flag\_sp} and \texttt{flag\_mg\_fe} are set to zero in order to remove stars with inaccurate stellar parameters and $\mgfe$ measurements. We follow the recommendation in \citet{buder2024} and do not apply the flag \texttt{flag\_fe\_h}, as \edit{up to 34\% of stars with detectable Fe lines have been flagged as non-detections because of an inappropriate choice of reference spectra}. Rapid rotators and lithium rich giants have anomalous spectra which are not caused by a star's evolutionary state. We remove these stars through restricting $\vsini<7\,\kms$ and $\rm{A(Li)}<1.5$\edit{, where we set \texttt{flag\_a\_li}$<2$ in order to remove stars with poor Li abundance detections}. To increase the spectral fidelity, we remove stars with $\snr$ per pixel below 10 in CCD1, $\snr$ below 20 in CCD2, and $\snr$ below 30 in CCD3 and CCD4. All observed spectra have been shifted to the rest frame and linearly interpolated onto the same wavelength grid.

\subsection{Asteroseismic catalogue}
We adopt asteroseismic observables and the stellar evolutionary state from the TESS \citep{ricker2014} and K2 \citep{howell2014} surveys. The frequency of maximum acoustic power ($\numax$) and the frequency separation between overtone modes ($\dnu$) are extracted from the power spectrum using the \textsc{SYD} pipeline \citep{huber2009, huber2011, yu2018}, which takes initial guess $\numax$ values from \citet{hon2018a}. The $\dnu$ values are then vetted through a neural network-based classifier \citep{reyes2022}\edit{, where we adopt the same vetting cut-off threshold of 0.5}. Lastly, the evolutionary state is classified using a convolutional neural network on the folded spectrum \citep{hon2017, hon2018}. \edit{Stars are classified in a continuum from 0 (a clear RGB star), to 1 (a clear RC star), with the vast majority of stars having values near 0 or 1. To avoid the ambiguous classifications near 0.5, we remove stars with classification between 0.3 to 0.7.}

\subsection{Crossmatched catalogue}
\begin{figure*}
    \centering
    \includegraphics[width=\linewidth]{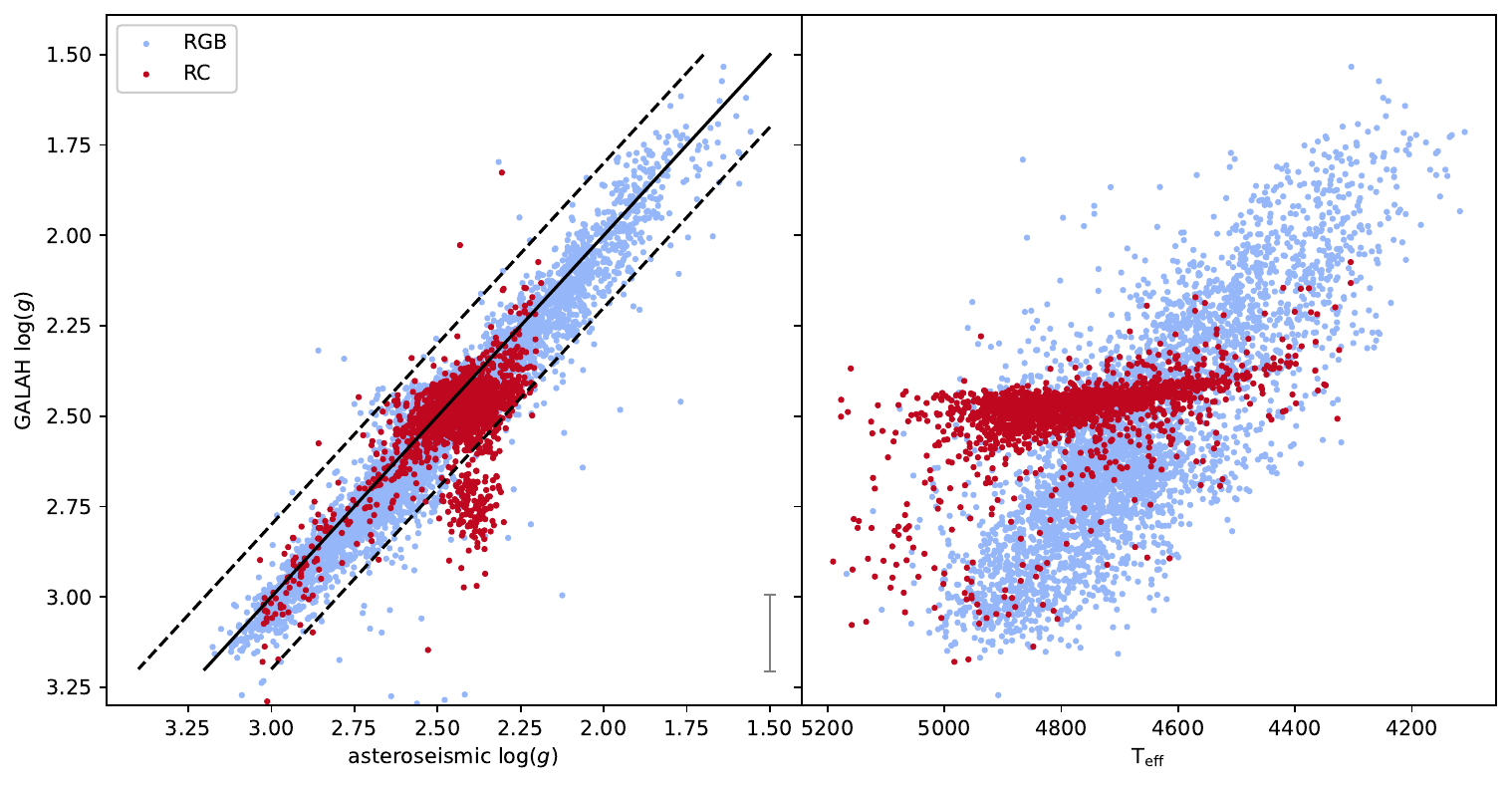}
    \caption{The left panel shows \galah $\logg$ with a representative errorbar compared to asteroseismic $\logg$. RGB stars are shown in blue whilst RC stars are shown in red. The solid black line indicates a one-to-one relationship between the two different $\logg$. The dashed black lines are offset by 0.2\,dex from the one-to-one relationship line, and stars with $\logg$ difference larger than 0.2 are removed from the crossmatched catalogue. The right panel shows the $\teff$ and $\logg$ space that the crossmatched stars occupy after removing stars with a $\logg$ discrepancy.}
    \label{fig:logg}
\end{figure*}

We combine the spectroscopic data with asteroseismic data by crossmatching between stars \edit{with the smallest angular distance} on sky. Using this crossmatched catalogue, we calculate asteroseismic ($\logg$):
\begin{equation}
    \logg = \log \left( \frac{\numax}{\nu_{\rm{max},\odot}} \right) + 0.5 \log \left( \frac{\teff}{\rm{T}_{\rm{eff},\odot}} \right) + \log(g_{\odot}),
\end{equation}
where $\teff$ is the effective temperature. We adopt Solar parameters $\rm{T}_{\rm{eff},\odot}=5772$\,K and $\log(g_{\odot})=4.438$ from \citet{prsa2016}, and $\nu_{\rm{max},\odot}=3090\,\muhz$ from \citet{huber2011}. Whilst \galah measures $\logg$ through its definition $(g \propto M/R^2)$ and the Stefan-Boltzmann scaling relation \citep{buder2024}. We show \galah $\logg$ compared to the asteroseismic $\logg$ in the left panel of Fig.~\ref{fig:logg}. \edit{In general, there is good agreement between asteroseismic and spectroscopic $\logg$, with the exception of some RC stars with \galah $\logg \approx 2.8$ and asteroseismic $\logg \approx 2.4$. These stars are overestimated in spectroscopic $\logg$ due to a mismatch of isochrones and stellar spectroscopic parameters, causing} primary RC stars to be misidentified as secondary RC stars \citep{buder2024}. 
\galah measures stellar parameters iteratively and consistently, in order to retain consistency, we adopt the \galah $\logg$. However, we remove stars with $\logg$ difference of more than 0.2\,dex \edit{in order to remove these misidentified secondary RC stars}. \edit{In total we crossmatch 8163 stars; after applying the aforementioned quality cuts, \total stars remain, out of which \rgb are RGB stars and \rc are RC stars.} These \edit{remaining} crossmatched stars are shown in the right panel of Fig.~\ref{fig:logg}.

Using \galah $\teff$, $\feh$ and asteroseismic $\numax$, $\dnu$, we calculate mass and radius using \asfgrid \citep{sharma2016, stello2022}. \asfgrid computes mass and radius using the asteroseismic scaling relations and Stefan–Boltzmann luminosity law in addition to a corrected $\dnu$ interpolated over a grid \citep{stello2022}. \edit{This correction is necessary to account for the well-known fact that $\dnu$ does not scale precisely with $\rho^{1/2}$ in red giant stars (see \citealt{sharma2016} for details)}. 

\section{Method}
\label{sec:method}
Stellar spectra contain information on effective temperature ($\teff$), surface gravity ($\logg$), metallicity ($\feh$), and elemental abundances. 
As a result, it is difficult to separate RGB and RC stars using spectroscopy alone because they occupy the same $\teff$, $\logg$, and $\feh$ space within uncertainties. 

In order to find spectroscopic differences caused by the evolutionary state, we need to control the observed spectroscopic parameters. We achieve this with a powerful differential technique by taking pairs of RGB and RC stars with similar spectroscopic parameters -- we refer to these as matched pairs. 
Such matched pairs of RGB and RC stars must have a difference in mass. \edit{This is because when a star evolves from the hydrogen-shell fusion state at the tip of the red giant branch to begin helium fusion, its outer envelope contracts making the star smaller and therefore hotter (i.e. as per the Stefan Boltzmann equation). If a RGB and a RC star have the same stellar parameters: $\teff$ and $\logg$ (which implies the stars have the same radius), then the RGB star must be less massive than the RC star.}  

Whilst mass cannot be directly measured from stellar spectra, secondary effects such as changes in carbon, nitrogen, and oxygen abundances due to \edit{the first }dredge up and deep mixing should be visible in the spectra of our matched pairs. \edit{At the base of the red giant branch, the convective envelope deepens into the core, dredging up material from the stellar core to the stellar surface \citep{iben1964, iben1967, lambert1981}. As a consequence, CNO-processed material in the core is brought to the surface of the star, modifying the observed $\cn$ and $\cisotope$ ratio \citep{lambert1977, lambert1981b, smith1985, halabi2015, masseron2015, martig2016, ness2016}. The amount of CNO-processed material in the core is a function of both stellar mass and the depth of the excursion of the convective envelope into the core. For example, lower mass stars have a lower fraction of core nitrogen abundance and a shallower excursion of the convection zone. Subsequently, lower mass stars at fixed metallicity have higher $\cn$. Deep mixing further modifies $\cn$ and $\cisotope$ ratio in a similar direction in stars beyond the RGB bump \citep{lagarde2012}. }

\subsection{Matched pairs}
We define a matched pair as two stars with different evolutionary states but similar spectroscopic and observational parameters: $\delta\teff<50$\,K, $\delta\logg<0.15$, $\delta\feh<0.05$\,dex, $\delta\mgfe<0.05$\,dex, and $\left|\frac{(\snr)_{\rm{RC}}}{(\snr)_{\rm{RGB}}}-1\right| <0.2$ for all 4 CCDs of the HERMES instrument. These ranges are approximately based on \galah uncertainties. These parameter criteria ensure similar spectroscopic properties and metallicity contribution from core collapse and thermonuclear supernovae \citep{sayeed2024}. The $\snr$ criteria ensure similar observational properties, creating RGB and RC pairs which have spectral differences driven by the remaining small parameter differences inherited from their respective evolutionary states. For each RC star, we randomly pick a RGB star that satisfies the aforementioned criteria, resulting in a total of $n=\pairs$ matched pairs. \edit{We note that unique RGB stars can be present in more than one matched pair, this is further discussed in Appendix~\ref{app:unique}.}

\begin{figure*}
    \centering
    \includegraphics[width=0.8\linewidth]{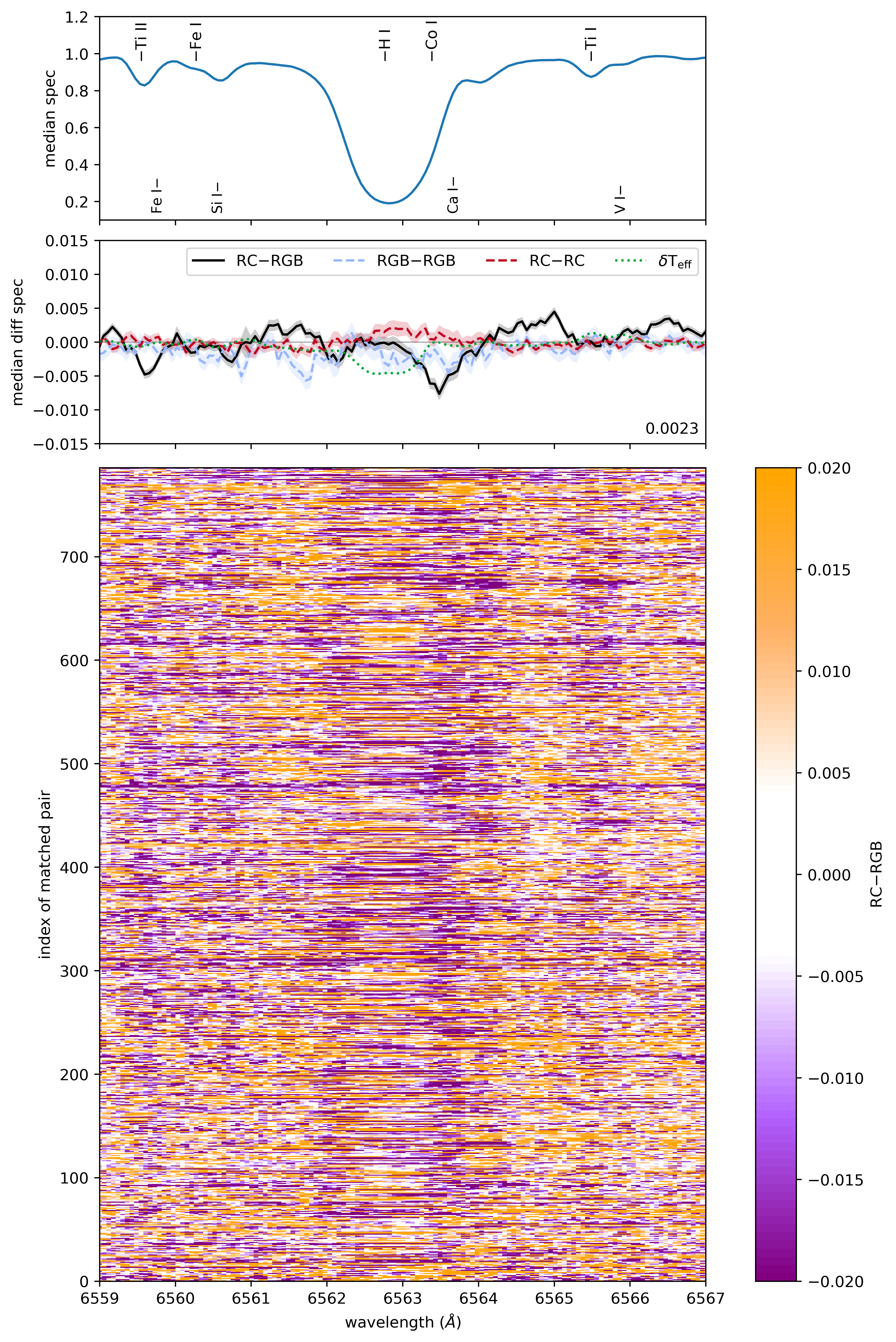}
    \caption{The top panel is the median spectrum for RGB stars \edit{with labelled atomic spectral features}, to show the spectral features in this wavelength region. The middle panel shows the median difference spectra: RC$-$RGB stars (solid black), RGB$-$RGB stars (dashed blue), and RC$-$RC stars (dashed red). The standard error on the difference spectra (Eq.~\ref{eq:std}) is shown for all 3 difference spectra in the corresponding colours. In order to show the deviation of the RC$-$RGB median difference spectrum from zero, we plot a \edit{thin gray line} at zero, and report the standard deviation for the \ha wavelength region on the bottom right. The difference spectrum for a change of 10\,K in $\teff$ is calculated through \pysme and shown in dotted green. The bottom panel shows the difference spectra for all RC$-$RGB matched pairs, with a truncated colorbar.}
    \label{fig:waterfall}
\end{figure*}

For each matched pair, we calculate a difference spectrum, $\delta f$, given by:
\begin{equation}
    \delta f_{i} = f_{{\rm RC},i} - f_{{\rm RGB},i},
\end{equation}
where $i\in \{ 1,\ldots,n=\pairs \}$ is the index of the matched pair, $f_{\rm{RC}}$ is the normalised flux of the RC star, and $f_{\rm{RGB}}$ is the normalised flux of the RGB star. The difference spectra on the \ha region for all matched pairs is shown in the bottom panel of Fig.~\ref{fig:waterfall}. 

The median RGB spectrum, $\rm{med}(f_{\rm{RGB}})$, is given by:
\begin{equation}
    {\rm med}(f_{\rm{RGB}}) 
    =
    {\rm med}(f_{{\rm RGB}, 1}, \ldots, f_{{\rm RGB}, n}).
\end{equation}
This median RGB spectrum is shown in the top panel of Fig.~\ref{fig:waterfall} as a reference for the typical spectral lines present in the \ha wavelength region. The median RC spectrum is not visibly different from the median RGB spectrum and is therefore not shown.

We want to study the difference in spectra caused by evolutionary state, not individual matched pairs, so we calculate the median difference spectrum, $\Delta f$, defined as:
\begin{equation}
    \Delta f = {\rm med}( \delta f_{1}, \ldots, \delta f_{n} ).
\end{equation}
The median difference spectrum for the \ha region is shown in the middle panel of Fig.~\ref{fig:waterfall}. The standard deviation of $\Delta f$ is a representation of how much $\Delta f$ deviates from zero, and is also shown in the middle panel of Fig.~\ref{fig:waterfall}. For the \ha region, this is 0.2\%, which is a small difference. Therefore, the rest of the methodology investigates the significance of this difference.

\subsection{Control pairs}
To quantify the effect of the matched methodology on the median difference spectrum, we find two additional sets of matched stars. We take the RGB stars from the existing matched pairs, and find their matched RGB stars using the same criteria. This gives \rgbpairs RGB to RGB matched pairs. We repeat this process for RC stars, and find \rcpairs RC to RC matched pairs. The median difference spectra for the RGB$-$RGB matches and RC$-$RC matches are shown in the middle panel of Fig.~\ref{fig:waterfall}. These additional difference spectra are smaller compared to the RC$-$RGB matched set, indicating that the difference in RC and RGB spectra is not caused by our matched pair selection. Instead, some of this difference must be due to their different evolutionary states.

\begin{figure*}
    \centering
    \includegraphics[width=0.49\linewidth]{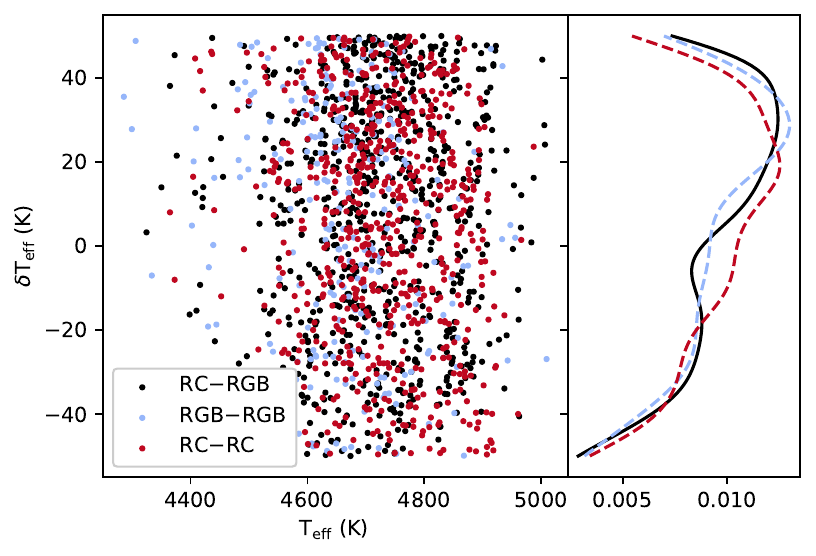}
    \includegraphics[width=0.49\linewidth]{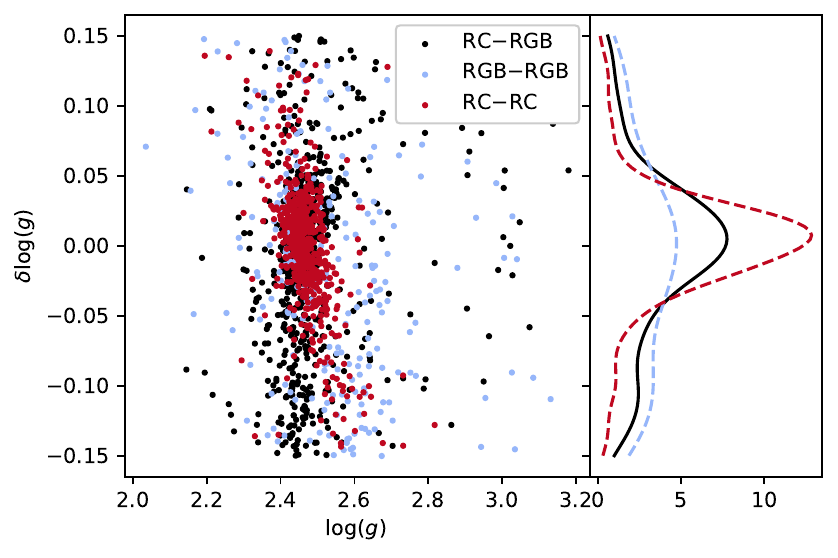}
    \includegraphics[width=0.49\linewidth]{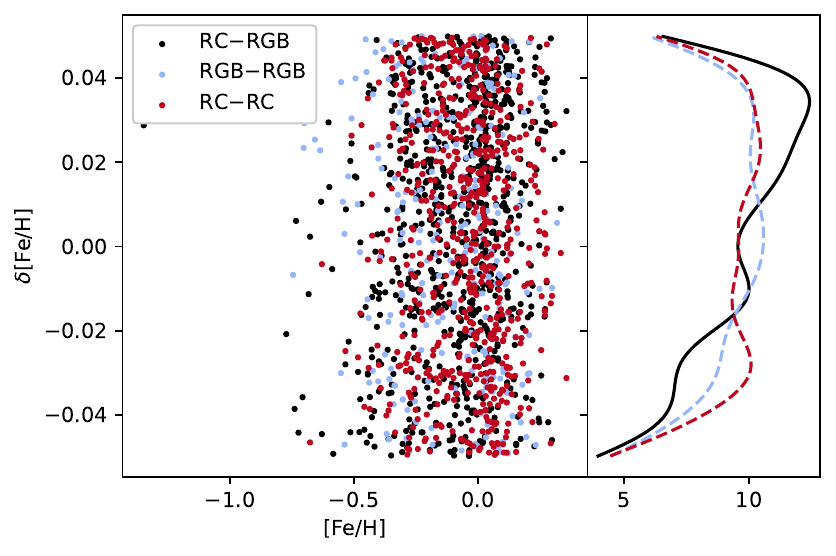}
    \includegraphics[width=0.49\linewidth]{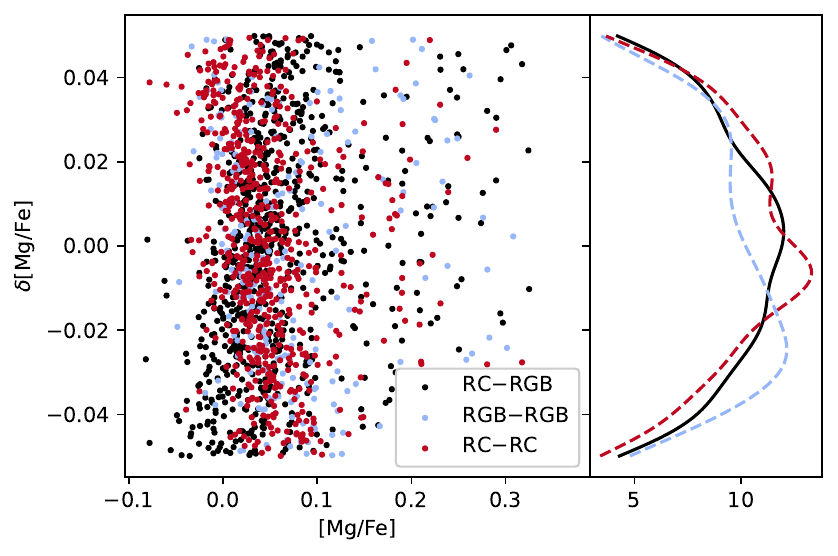}
    \caption{The correlation between RC stellar parameters and the stellar parameter difference, shown in the left panel of each subfigure; and the distribution of stellar parameter difference modelled using a kernel density estimator, shown in the right panel of each subfigure. Each subfigure shows a different stellar parameter: $\teff$ and $\logg$ in the top row, $\feh$ and $\mgfe$ in the bottom row. All three matched sets are shown: RC$-$RGB (solid black), RGB$-$RGB (dashed blue), and RC$-$RC (dashed red).}
    \label{fig:param_hist}
\end{figure*}

Despite selecting matched pairs that have similar stellar parameters, there is still an expected overall non-zero difference in stellar parameters between matched pairs. Table~\ref{tab:delta_sp} shows the median and mean difference in stellar parameters for all three matched sets. This non-zero mean difference further implies that the distribution in stellar parameter difference is non-uniform, as shown in Fig.~\ref{fig:param_hist}. Despite the non-zero and non-uniform distribution of the difference in stellar parameters, it cannot fully account for the signal in the RC$-$RGB median difference spectrum that we see. The distribution, median difference, and mean difference in stellar parameters are very similar between the three matched sets. Therefore, we expect the impact on the median difference spectrum to be similar. However, the RC$-$RGB matched median difference spectrum shows a larger signal than the RC$-$RC and RGB$-$RGB matched median difference spectrum, implying that some of the RC$-$RGB median difference spectrum is caused by evolutionary state parameters that are not included in our matched criteria.

Furthermore, we use \pysme \citep{Wehrhahn2023} to calculate synthetic spectra at 10\,K apart, and show that the median difference spectrum cannot be fully explained by the non-zero $\delta \teff$ \edit{reported in Table~\ref{tab:delta_sp}}. The \pysme difference spectrum is shown in the middle panel of Fig.~\ref{fig:waterfall}. However, we caution that spectral synthesis of the \ha line in giant stars differs from the observed spectra by up to 20\%--a well known issue \citep{bergemann2016}--therefore, the flux difference that a 10\,K $\teff$ perturbation creates could be different in reality. We further explore the impact of selection effects in the matched methodology on the median difference spectrum in Appendix~\ref{app:impact}.

\begin{table*}
    \centering
    \caption{The median and mean difference in stellar parameters for the full crossmatched dataset and the matched sets. These differences are of a similar magnitude, and so the impact of stellar parameter on the median difference spectra is similar between the matched sets.}
    \begin{tabular}{c|ccccccccc}
\hline
Parameter & Units & \multicolumn{2}{c}{Full} & \multicolumn{2}{c}{RC$-$RGB} & \multicolumn{2}{c}{RGB$-$RGB} & \multicolumn{2}{c}{RC$-$RC} \\
& & Median & Mean & Median & Mean & Median & Mean & Median & Mean \\
\hline 
\hline
$\teff$ & K & $87.04$ & $101.31$ & $10.87$ & $7.17$ & $12.12$ & $7.23$ & $7.96$ & $4.89$ \\
$\logg\ /\ 10^{-2}$ & $\log(\cms)$ & $-11$ & $-4.7$ & $-0.19$ & $-0.75$ & $-1.4$ & $-1.4$ & $0.32$ & $-0.0047$ \\
$\feh\ /\ 10^{-2}$ & dex & $15$ & $13$ & $0.92$ & $0.58$ & $0.32$ & $0.26$ & $0.25$ & $0.22$ \\
$\mgfe\ /\ 10^{-3}$ & dex & $-52$ & $-53$ & $-0.25$ & $-0.35$ & $-5.6$ & $-3.1$ & $0.38$ & $0.94$ \\
\hline
    \end{tabular}
    \label{tab:delta_sp}
\end{table*}

\subsection{Flux uncertainties}

The median difference spectrum is more significant than its standard error and uncertainty. The standard error of the median difference spectra, ${\rm std}_{\Delta f}$, is approximated by:
\begin{equation}
    {\rm std}_{\Delta f} = \frac{{\rm std}(\delta f_1, \ldots, \delta f_n)}{\sqrt{n}}.
    \label{eq:std}
\end{equation}
The uncertainty of the median difference spectrum assuming a Gaussian distribution, $\sigma_{\Delta f}$, is approximated by:
\begin{align}
    \sigma_{\Delta f} = \left( \sum_{i=1}^{n}\frac{1}{\sigma_{{\rm RGB},i}^2 + \sigma_{{\rm RC},i}^2} \right)^{-\frac{1}{2}},
    \label{eq:err}
\end{align}
where $\sigma_{\rm{RGB}}$ is the flux error of the RGB star and $\sigma_{\rm{RC}}$ is the flux error of the RC star. See Appendix~\ref{app:stats} for a derivation of Eq.~\ref{eq:err}. $\sigma_{\Delta f}$ is smaller than ${\rm std}_{\Delta f}$ by a factor of 1.5 in the \ha region, 
therefore, we show the larger error, ${\rm std}_{\Delta f}$, from Eq.~\ref{eq:std} in the middle panel of Fig.~\ref{fig:waterfall} as shaded regions for all three matched sets. Despite the sub-percent magnitude of the median difference spectrum, the RC$-$RGB median difference spectrum is significant when compared to the uncertainty estimates. 

\section{Results}
\label{sec:results}

The median difference spectrum for RC$-$RGB matched stars is non-zero and larger than the flux error and control matched sets for many wavelengths within \galah spectra. If additional parameters of stellar evolutionary state not included in our criteria (i.e., mass) do not affect spectra, then we expect the median difference spectrum to be zero, as the matched pairs were selected to have similar stellar parameters. Therefore, the spectroscopic signature of evolutionary state is where the median difference spectrum is non-zero. We find that the RC$-$RGB matched median difference spectrum deviates from zero where $\ctwo$ and CN bands are located, and in the line-width near the inner wings of the \ha and \hb lines.

\subsection{\texorpdfstring{$\ctwo$}{C2} and CN}
\begin{figure*}
    \centering
    \includegraphics[width=\linewidth]{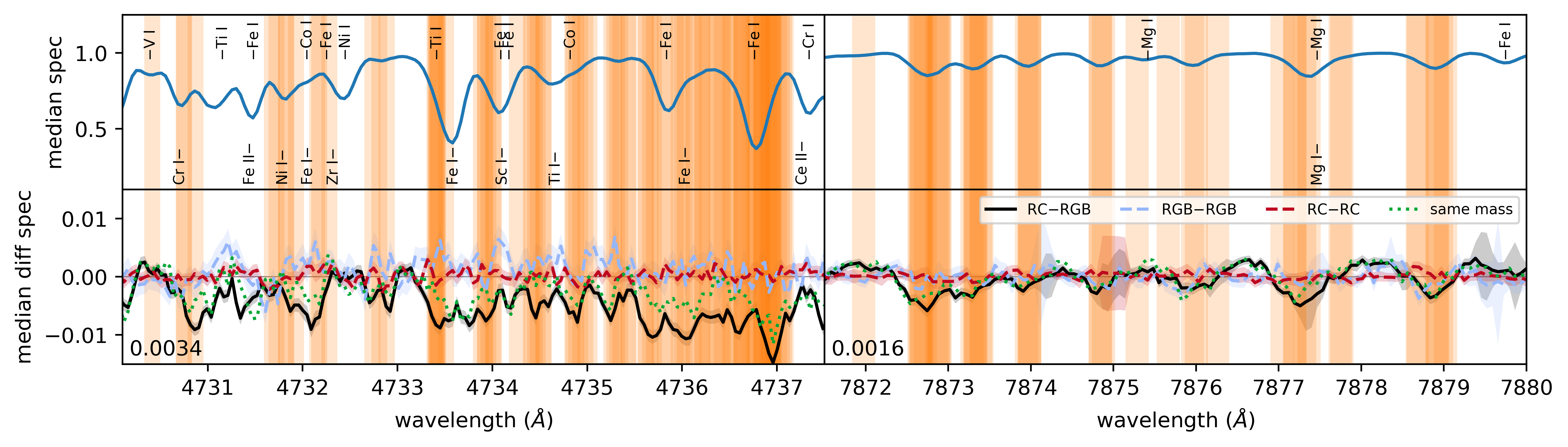}
    \caption{The median difference spectrum is correlated with the Swan bands (left) and the CN bands (right). For each subfigure, the top panel shows the median RGB spectrum \edit{with labelled atomic spectral features}, as a reference for the spectral features of this wavelength region. The bottom panel shows the median difference spectrum for RC$-$RGB stars (solid black); the uncertainty in the median difference spectrum given by Eq.~\ref{eq:err} (shaded gray); RGB$-$RGB matched pairs (dashed blue) and RC$-$RC matched pairs (dashed red). $\ctwo$ and CN molecular lines are shown in both panels as orange shaded rectangles, where darker orange indicates overlapping $\ctwo$ and CN lines. The median difference spectrum for the mass constrained matched set is shown in the bottom panel (dotted green) of each subfigure and is discussed in Section~\ref{sec:discussion}.}
    \label{fig:CN}
\end{figure*}

We find that the spectroscopic signature of evolutionary state correlates with $\ctwo$ and CN molecular band heads. Fig.~\ref{fig:CN} shows the Swan bands in CCD1 and CN bands in CCD4 captured by \galah. We extract a stellar line list at $\teff=4750$\,K, $\logg=2.5$, microturbulence $\vmic=1.46$\,$\kms$\ from VALD3 \citep{ryabchikova2015}, and show $\ctwo$ and CN molecular lines \citep{OK, LWHS, PTP} with VALD3 central depth parameter larger than 0.005 shaded in orange. The width of the shaded region ($w$) is indicative of the instrumental broadening of the molecular line, and is defined as $w=\lambda_l/R$, where $\lambda_l$ is the wavelength of the molecular line, and $R$ is the resolution of the spectra. The median difference spectrum deviates from zero at the same wavelengths as the $\ctwo$ and CN molecular band heads. The median difference spectrum is negative, implying that the $\ctwo$ and CN features are stronger in RGB stars than RC stars. We note that despite $\ctwo$ and CN molecular lines being present at many other wavelengths, the correlation with the median difference spectrum is only clear in the band heads. Band heads are where many lines from the same molecular species are present; therefore, this correlation only being clear in the band heads implies that every molecular line carries a small bit of evolutionary state signal which only becomes visible when many of these lines are close in wavelength. 

\subsection{Line-width}
\begin{figure*}
    \centering
    \includegraphics[width=\linewidth]{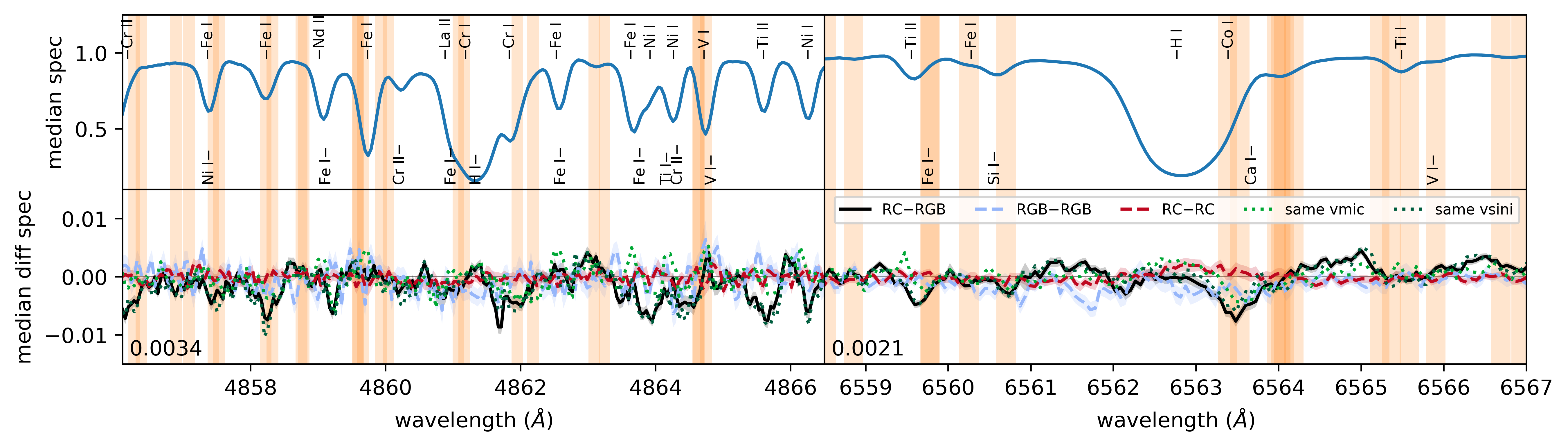}
    \caption{
    The line-width of the \ha (right) and \hb (left) line is different between RC and RGB stars of similar stellar parameters. Labels follow Fig.~\ref{fig:CN}. The median difference spectrum for matched sets matched to agree in $\vmic$ and $\vsini$ respectively, in addition to $\teff$, $\logg$, $\feh$, and $\mgfe$, are shown in the bottom panel (dotted \edit{green} lines) of each subfigure and are discussed in Section~\ref{sec:discussion}.}
    \label{fig:vmic}
\end{figure*}

We see a difference in the line-width of the \edit{\ha and \hb lines} in the spectrum, shown in Fig.~\ref{fig:vmic}. The median difference spectrum is close to zero in the core of the line, but negative in the wings of the line, implying that lines in RC stars are broader than in RGB stars. Whilst we only show \ha and \hb in Fig.~\ref{fig:vmic}, the line-width difference can be seen from other lines in the spectrum. However, it becomes difficult to distinguish this difference in line-width when the lines are heavily blended. The \ha median difference spectrum is visibly asymmetric, likely caused by both a difference in the line-width and two CN molecular lines situated in the red wing of the \ha line at $\lambda \approx 6563.4$\,\AA. Whilst the smaller negative median difference spectrum in the blue wing of the \ha line is caused only by a difference in the line-width.

\subsection{Formation depth}
\label{sec:formation_depth}
\begin{figure}
    \centering
    \includegraphics[width=\linewidth]{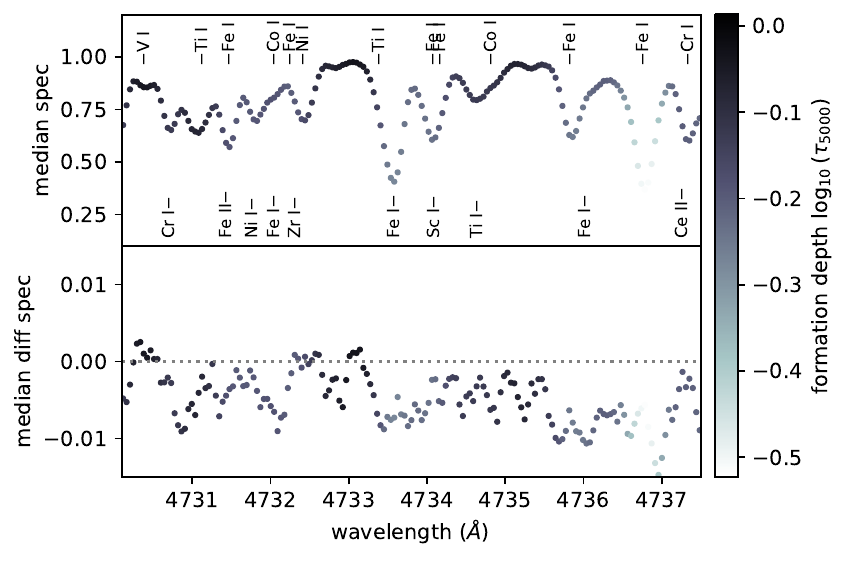}
    \caption{\edit{The top panel shows the median RGB spectrum with labelled atomic spectral features coloured by formation depth.} Where normalised flux is closer to 1, the formation depth is geometrically higher, and vice versa. \edit{The bottom panel shows the median difference spectrum coloured by formation depth.} There is no correlation between the formation depth and the median difference spectrum.}
    \label{fig:formation_depth}
\end{figure}

\edit{The median difference spectrum does not appear to be driven by the formation depth}. We define the formation depth per wavelength as the geometric depth at which the majority of photons for a wavelength originate. We approximate the geometric depth using the optical depth ($\tau_{\lambda}$). Specifically, we calculate $\tau_{5000}$ where $\tau_{\lambda}=1$ for every wavelength using Korg \citep{wheeler2023}, and adopt $\log_{10}(\tau_{5000})$ as our proxy for formation depth. We compare the formation depth to the median RGB spectrum and median difference spectrum in Fig.~\ref{fig:formation_depth}. We find that the formation depth is higher where the normalised flux is closer to zero, e.g. at $\lambda = 4736.8$\,\AA, as expected. However, we do not see a correlation with the amplitude of the median difference spectrum, indicating that the origin of the spectroscopic signal of evolutionary state is not related to the geometric height of the atmosphere. 

\section{Discussion}
\label{sec:discussion}

We find spectroscopic signatures of evolutionary state in the $\ctwo$ and CN bands, and in the wings of the \ha and \hb lines. In this section, we relate these results to physical parameters. 

\subsection{\texorpdfstring{$\ctwo$}{C2} and CN}
\label{sec:discussion1}
For a pair of RGB and RC stars to have similar spectroscopic stellar parameters, they must have different masses. We find that the average mass of our RGB matched stars is $1.39\,\msun$ with standard deviation $0.35\,\msun$, while for our RC matched stars it is $1.14\,\msun$ with standard deviation $0.33\,\msun$. To further investigate the impact of mass on our median difference spectrum, we make an additional matched set, using the same matched criteria as previously, except with an extra mass restriction of $\delta\mass<0.3\,\msun$. The median difference spectrum of this mass restricted matched set is shown in the bottom panel of Fig.~\ref{fig:CN}. The amplitude of the mass restricted median difference spectrum is smaller than the amplitude of the median difference spectrum in the $\ctwo$ and CN bands, indicating that part of the evolutionary state signal in these regions is due to a difference in mass. 

The strength of the $\ctwo$ and CN lines depends on the molecular equilibrium and isotopic ratio of relevant species. The CNO abundances in giant stars have been impacted by the first dredge up, and RC CNO abundances have been further modified through deep mixing. Similarly, C isotopic ratios are modified through both the first dredge up and deep mixing. Furthermore, the amount that these abundances and isotopic ratios change by depends on mass and metallicity. In Table~\ref{tab:lagarde} we report CNO abundances and isotopic ratios at the average mass and $\logg$ of our RGB and RC stars, predicted using stellar evolution models incorporating thermohaline instability and rotation-induced mixing \citep{lagarde2012}.

\begin{table}
    \centering
    \caption{The predicted CNO abundance and C isotopic ratios for stellar evolution models representing the mean mass of RGB and RC stars in our RC$-$RGB matched pairs. We adopt stellar evolution models from \citet{lagarde2012} at $\logg=2.5$ and Solar metallicity (${\rm Z}=0.014$), these models incorporate thermohaline instability and rotation-induced mixing. These values are linearly interpolated from grid models.}
    \begin{tabular}{cccc}
        \hline
         & Units & RGB & RC \\
        \hline 
        \hline 
        $\mass$ & $\msun$ & $1.39$ & $1.14 $ \\
        A(C) & dex & $8.286$ & $8.283 $ \\
        A(N) & dex & $8.187$ & $8.193 $ \\
        A(O) & dex & $8.710$ & $8.714 $ \\
        ${}^{12}{\rm C}:{}^{13}{\rm C}$ & - & $95:5$ & $89:11$ \\
        \hline
    \end{tabular}
    \label{tab:lagarde}
\end{table}

\begin{figure*}
    \centering
    \includegraphics[width=\linewidth]{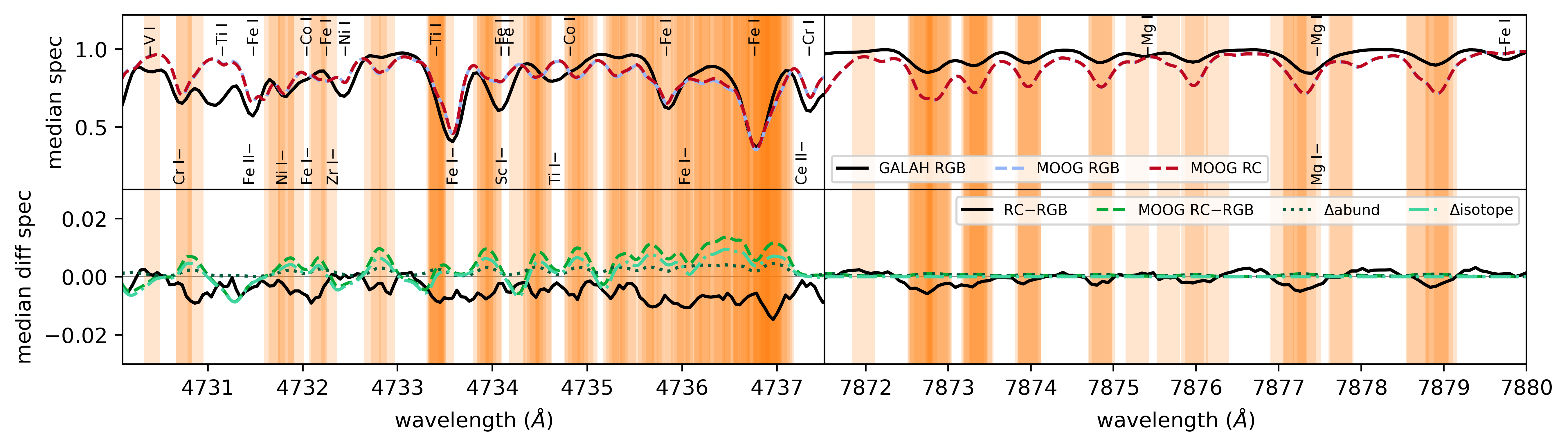}
    \caption{The top panel shows the normalised flux for \galah observed median RGB spectrum (black)  \edit{with labelled atomic spectral features}, \moog synthetic spectrum calculated with RGB abundances and isotopic ratio (blue), and RC abundances and isotopic ratio (red) based on stellar evolution models \citep{lagarde2012}. Bottom panel shows difference spectra: observed RC$-$RGB matched pairs (black), \moog synthetic RC$-$RGB using \citet{lagarde2012} abundances and isotopic ratios (dashed green), \moog synthetic different abundances same isotopic ratios (dotted dark green), and \moog synthetic same abundances different isotopic ratios (dash dotted light green).}
    \label{fig:isotope}
\end{figure*}

To investigate the impact of the first dredge up and deep mixing on the median difference spectrum, we compute synthetic spectra using the 1D local thermodynamic equilibrium spectrum synthesis code \moog \citep{Sneden1973}. We adopt the average $\teff$, $\logg$, $\feh$, and $\mgfe$ of the RC$-$RGB matched set, along with CNO abundances and isotopic ratios from \citet{lagarde2012} as listed in Table~\ref{tab:lagarde}, and apply Solar abundances for all other elements.
We run the 2017 version of \moog\footnote{\url{https://github.com/alexji/moog17scat}} with the \pymoogi GUI \citep{Adamow2017} using \atlas model atmospheres interpolated using \pykmod\footnote{\url{https://github.com/kolecki4/PyKMOD}} and a line list generated by linemake \citep{Placco2021}. This line list adopts $\ctwo$ data from \cite{Ram+2014}, CN data from \cite{Sneden+2014} and CH data from \cite{Masseron+2014}. We show two \moog synthetic spectra in the top panel of Fig.~\ref{fig:isotope}: RGB abundances and isotopic ratio, and RC abundances and isotopic ratio. The difference between the \moog RC spectrum and RGB spectrum then corresponds to our observed RC$-$RGB median difference spectrum and is shown in the bottom panel of Fig.~\ref{fig:isotope}.
\edit{In both the Swan and CN bands,} the synthetic difference spectrum is \edit{positive}, despite the observed negative median difference spectrum. \edit{The magnitude of the \moog difference spectrum is roughly the same as the observed median difference spectrum in the Swan band. However, in the CN band, the magnitude of the \moog difference spectrum is smaller compared to the observed median difference spectrum.}

To determine whether the median difference spectrum depends on the change in CNO abundances or C isotopic ratio, we compute two additional synthetic spectra with \moog: RGB abundances with RC isotopic ratio, and RC abundances with RGB isotopic ratio. The synthetic difference spectrum with RC abundances but different isotopic ratios, and the difference spectrum with the RC isotopic ratio but different abundances, are shown in the bottom panel of Fig.~\ref{fig:isotope}. 
In both wavelength regions, the \moog RC$-$RGB difference spectrum \edit{is approximately the sum of both the change in abundance and the change in isotopic ratio. }
In the Swan bands, the change in isotopic ratio \edit{is larger than the change in abundance}; whilst in the CN bands wavelength region, the change in isotopic ratio \edit{is smaller than the change in abundance. }

We caution that these synthetic spectra were computed using abundances from stellar evolution models \citep{lagarde2012}, and do not match \galah observations, which measure abundances using radiative transfer codes \citep{buder2024}. Notably, the synthetic spectra differ from the observed spectra by up to 20\% at many wavelengths, as shown in the top panel of Fig.~\ref{fig:isotope}. The difference spectra could change; if the synthetic spectra are fitted to the observed normalised flux instead. However, synthetic spectra are not accurate to the sub-percent level, and will introduce model errors into such fits. 

\subsection{Line-width}
Line-width is controlled by stellar rotation, $\vsini$, and velocity fields within the stellar atmosphere, which are approximated by $v_{\rm mac}$ and $\vmic$ within 1D model atmospheres. 
\galah measures $\vmic$ and $\vsini$, where $\vsini$ represents the line broadening from both $v_{\rm mac}$ and stellar rotation. In order to determine the parameter driving the line broadening observed, we make additional matched sets. The $\vsini$ constrained matched set incorporates an extra constraint of $\delta(\vsini) < 0.8$\,$\kms$, whilst the $\vmic$ constrained matched set incorporates an extra constraint of $\delta\vmic < 0.07$\,$\kms$. We find that the $\vmic$ constrained matched set has a median difference spectrum closer to zero whilst the $\vsini$ constrained matched set does not (shown in bottom panel of Fig.~\ref{fig:vmic}), implying that the difference in line-width we observe is due to $\vmic$ and not $\vsini$. Furthermore, the median difference spectrum is mostly negative for both \ha and \hb, and the integrated median difference spectrum is: $-0.23$\,\AA, $-0.12$\,\AA, $-0.08$\,\AA, $-0.05$\,\AA, respectively for the 4 CCDs of HERMES. $\vsini$ conserves the line strength whilst $\vmic$ does not, further implying that the difference in line-width is caused by $\vmic$ and not $\vsini$. In addition, we find that $\vmic$ improves random forest classifier accuracy (see Appendix~\ref{app:rf}), whilst $\vsini$ does not. 

Our RGB matched stars have the same $\logg$ but are more massive than RC matched stars, as a result, these RGB matched stars must also have a larger radius. We incorporate an extra constraint of $\delta R<0.6\,R_\odot$ and compute the median difference spectrum. We find that the median difference spectrum for this radius restricted matched set is similar to the RC$-$RGB matched set, indicating that radius is not the cause of the observed difference in line-width. 

\begin{figure}
    \centering
    \includegraphics[width=\linewidth]{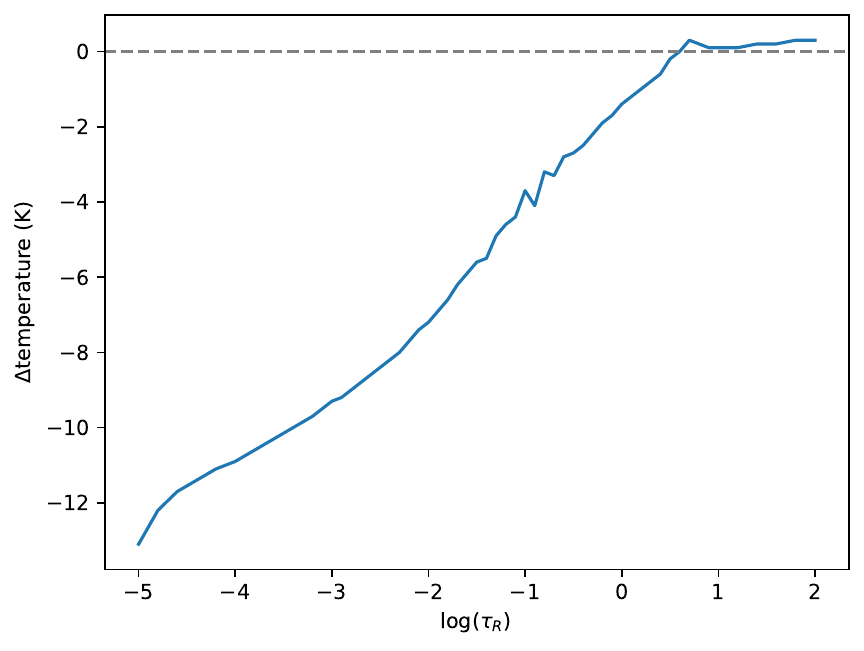}
    \caption{Temperature difference at different Rosseland optical depth for spherically symmetric stellar atmosphere models with $\mass=1\msun$ compared to $\mass=2\msun$. Model atmospheres from the MARCS grid \citep{gustafsson2008}, located at $\teff=5000$\,K, $\logg=2$, $\feh=0$, $\vmic=1$\,$\kms$.}
    \label{fig:marcs}
\end{figure}

$\vmic$ parameterises small scale velocity motions in the stellar photosphere. We measure an average $\vmic = 1.42\,\kms$ with $0.00014\,\kms$ standard error in the mean and $\sigma_{\vmic}=0.11\,\kms$ in our RGB matched stars, and an average $\vmic=1.49\,\kms$ with $0.00012\,\kms$ standard error in the mean and $\sigma_{\vmic}=0.098\,\kms$ in our RC matched stars. As previously discussed, our RC matched stars are lower mass than the corresponding RGB matched stars. Sphericity effects related to this mass difference cause the RC star's temperature to be lower by up to 13\,K in the line forming regions, as shown in Fig.~\ref{fig:marcs}. This lower temperature could cause RC stars to have weaker \ha wings than RGB stars. However, density and photospheric radius also change with mass, so the overall impact of mass on line strength is unclear. Although 3D hydrodynamic model atmospheres simulate convection from first principles and do not use the $\vmic$ parameter, RC and RGB stars are modelled in the same way. Therefore, there is currently no know difference in the velocity field of RC stars compared to RGB stars. It has been shown that changes in CNO abundances can affect the atmospheric structure \citep{gallagher2017, zhou2023}, however, no detailed studies have been conducted on the impact of CNO abundances on the velocity field in 3D stellar atmospheres.

The core of the \ha and \hb lines form in the chromosphere and are linked to stellar activity and age, where younger stars are more active. The average age of our RGB stars are 6.08 Gyr with standard deviation 2.23 Gyr, and 6.38 Gyr with standard deviation 2.04 Gyr for RC stars, estimated through isochrone fitting (see \citealt{buder2024} for details). Whilst we do not expect the majority of our stars to be active, there may be some younger RGB stars which have a more active chromosphere; studies have shown that young stars appear to exhibit elevated $\vmic$ values \citep[e.g.][]{baratella2020}. However, this is the opposite of what we observe in our matched set. We note that the majority of our matched RC stars are primary RC stars; using a $2\,\msun$ cut-off \citep{girardi1999}, only 16 pairs contain a secondary RC star. Therefore, the differences that we see between RC and RGB stars are mainly driven by primary RC stars. 

In addition to $\vmic$, mass also impacts the median difference spectrum in the red wing of the \ha line. \citet{bergemann2016} showed that the line-width of the \ha line is broader for lower mass stars. We find a similar result, where our RC stars have broader \ha and \hb lines along with lower mass compared to our RGB stars. This mass dependence is likely due to the CN molecular lines present in the red wing of the \ha line, where the strength of the CN molecular lines is related to mass (see Section~\ref{sec:discussion1}). In particular, the median difference spectrum in the red wing goes closer to zero when mass is restricted, reducing the asymmetry. 

\section{Conclusions}
\label{sec:conclusion}

In this work, we present a model-free analysis of optical spectroscopic evolutionary state signals, and find sub-percent amplitude changes at all wavelengths in the spectra caused by evolutionary state. Despite the small signal, we show that this is significant compared to control sets and uncertainty estimates. For RC and RGB stars at similar stellar parameters, we find stronger $\ctwo$ and CN features in RGB stars compared to RC stars, caused by a difference in stellar evolution and mass; and we find \edit{that the \ha and \hb lines are broader} in RC stars than RGB stars, caused by a difference in microturbulence, as measured by \galah. We find that rotational velocity, radius, and formation depth do not impact spectroscopic evolutionary state signals. This analysis is demonstrative of the utility of large surveys to undertake a data-driven and model-independent population analysis, revealing the subtle differences between RC and RGB stars in the line forming regions of the stellar atmosphere. Future studies classifying stellar evolutionary states for spectroscopic surveys can use these results to inform which parameters and wavelengths contain evolutionary state information. Given the small amplitudes of these residual differences, analysis of these in individual stars would require $\textrm{S/N} > 200$, at resolution $\approx 28000$.

Whilst we identify some of the sub-percent amplitude changes and link them to evolutionary state, there are still many wavelength regions which we did not study. Future work could look at these other wavelengths and find additional parameters that these amplitude differences depend on, and extend this model-free technique to infrared spectra where there are more carbon-related molecular lines. 

\section*{Acknowledgements}
We thank Adam Wheeler for providing formation depths for Section~\ref{sec:formation_depth}. We thank Alexander Soen for various discussions regarding Eq.~\ref{eq:err} and helping formulate Appendix~\ref{app:stats}.
This work has made use of the VALD database, operated at Uppsala University, the Institute of Astronomy RAS in Moscow, and the University of Vienna. 
E.W. and K.L. acknowledge funds from the European Research Council under the European Union’s Horizon 2020 research and innovation program (grant agreement 852977). K.L. and T.N. acknowledge support from the Knut and Alice Wallenberg Foundation.
D.S. is supported by the Australian Research Council (DP190100666).
X.F. acknowledges National Natural Science Foundation of China (NSFC) No. 12203100, the China Manned Space Project with No. CMS-CSST-2021-A08. 
PBD is supported by the Australian Government International Training Program (iRTP) Scholarship.

\section*{Data Availability}
The \galah DR4 spectroscopic and catalogue data are available on \url{https://www.galah-survey.org/dr4/overview/}. Other data available upon request. 

\bibliographystyle{mnras}
\bibliography{bibfile} 

\appendix

\section{Impact on median difference spectrum}
\label{app:impact}

The amplitude of the median difference spectrum is small but significant, as shown through comparison to controlled matched sets of RGB$-$RGB and RC$-$RC stars (see Section~\ref{sec:method}). Given the small amplitude, minor changes in the RC$-$RGB matched set could cause visible changes in the median difference spectrum. In this section, we investigate the impact of RC$-$RGB matched set definition on the median difference spectrum.

\subsection{Duplicate RGB stars} 
\label{app:unique}
\begin{figure}
    \centering
    \includegraphics[width=\linewidth]{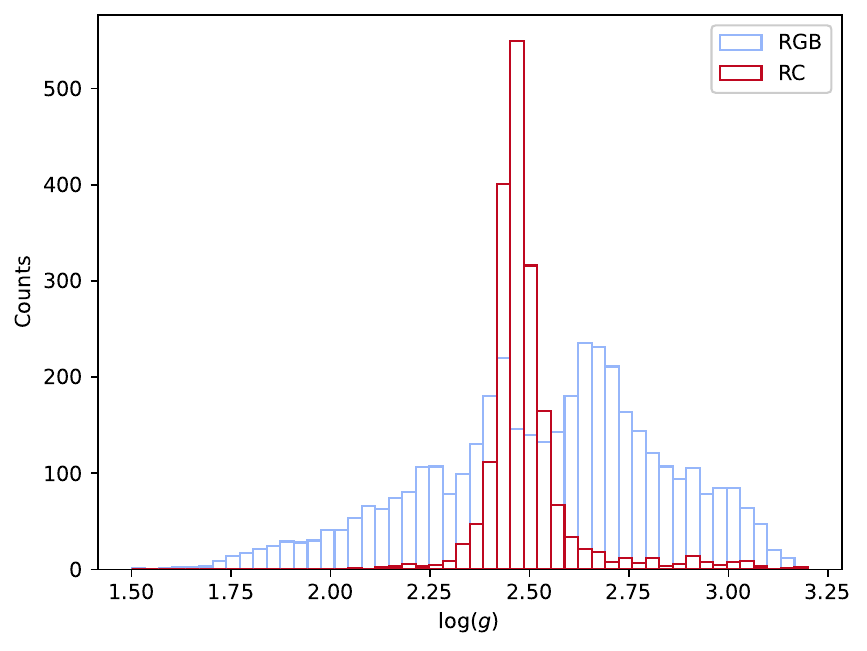}
    \caption{$\logg$ vs counts for the crossmatched catalogue, where RGB stars are shown in blue and RC stars are shown in red. The RC stars are more clustered in $\logg$ whilst RGB stars are more spread out.}
    \label{fig:logg_dist}
\end{figure}

There are repeated RGB stars in the RC$-$RGB matched set. For each RC star, we find a corresponding RGB star; however, the RGB stars are spread out in $\logg$ whilst RC stars are concentrated at $\logg \approx 2.5$, as shown in Fig.~\ref{fig:logg_dist}. As a result, the RGB matched stars are not unique. 

\begin{figure*}
    \centering
    \includegraphics[width=\linewidth]{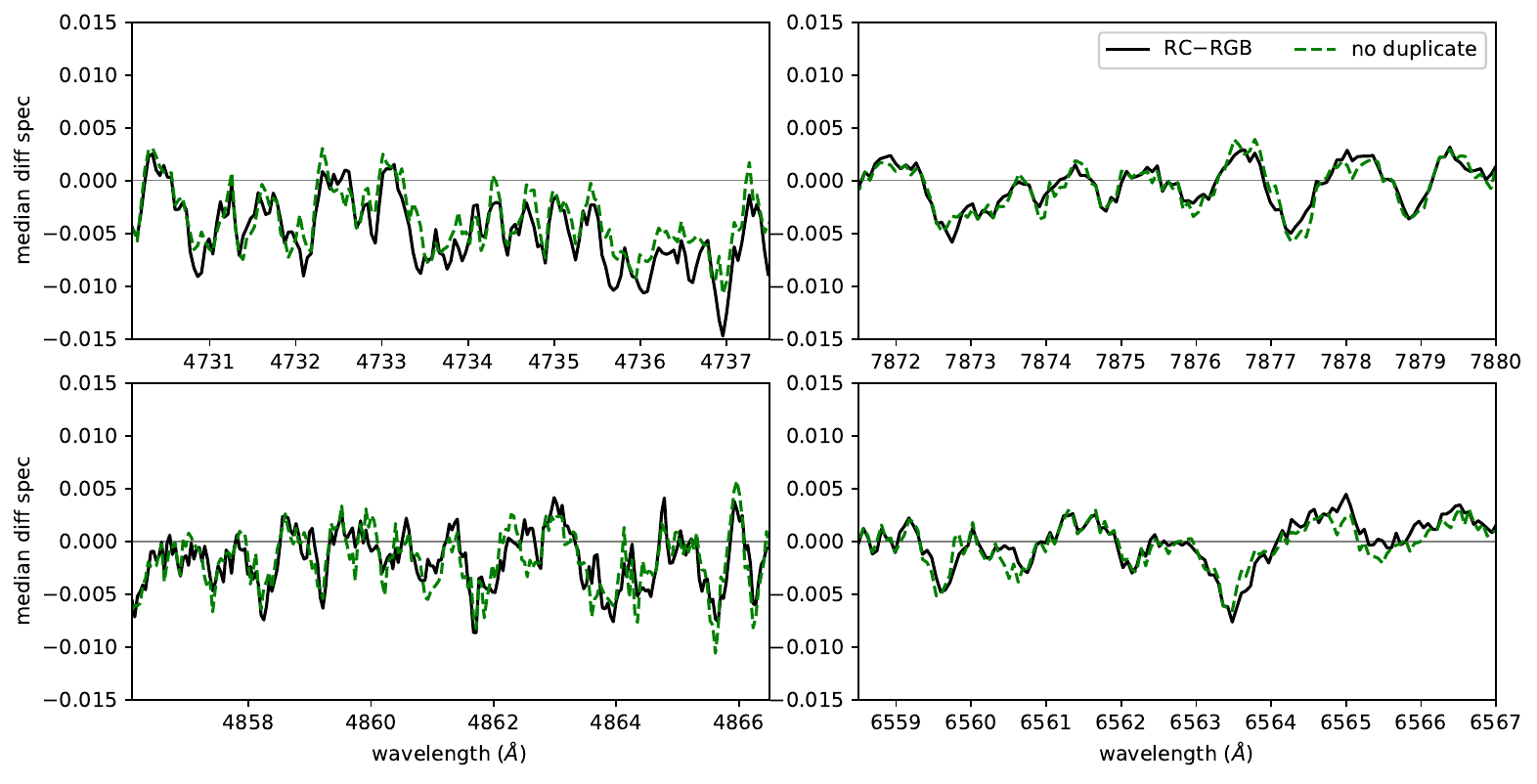}
    \caption{The median difference spectrum for the full RC$-$RGB matched set with duplicates (black) and a no duplicate matched set (green). The difference between the median different spectra is minimal, indicating that duplicate RGB stars do not affect the median difference spectrum.}
    \label{fig:dup}
\end{figure*}

We create a new matched set, where pairs containing a duplicate RGB star are removed, resulting in 394 matched pairs. The median difference spectrum for this no duplicate matched set is shown in Fig.~\ref{fig:dup}. The median difference spectrum is similar between the RC$-$RGB matched set and the no duplicate matched set, indicating that these duplicate RGB stars do not impact the spectroscopic signal of evolutionary states.

\subsection{Stellar parameters}
\begin{figure*}
    \centering
    \includegraphics[width=0.49\linewidth]{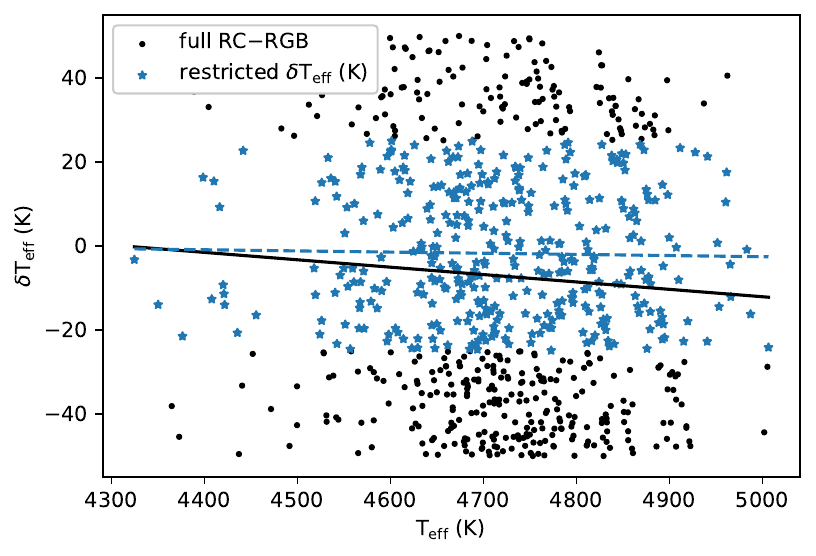}
    \includegraphics[width=0.49\linewidth]{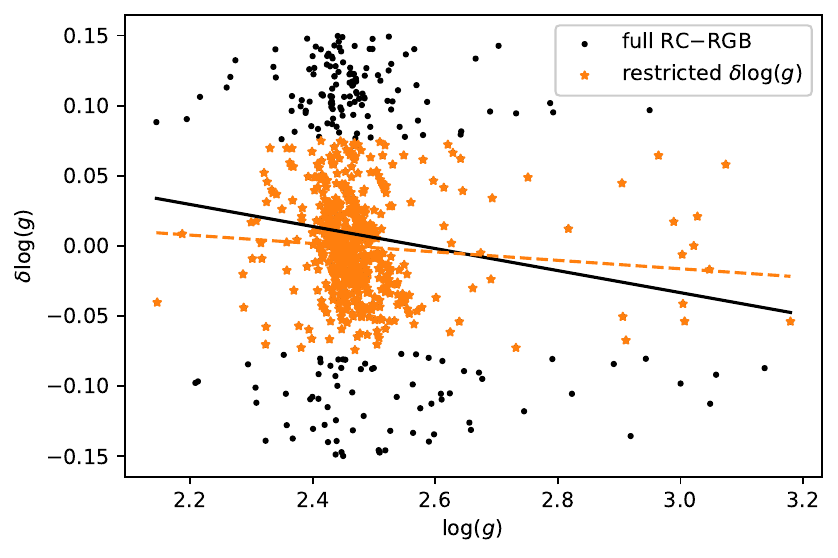}
    \includegraphics[width=0.49\linewidth]{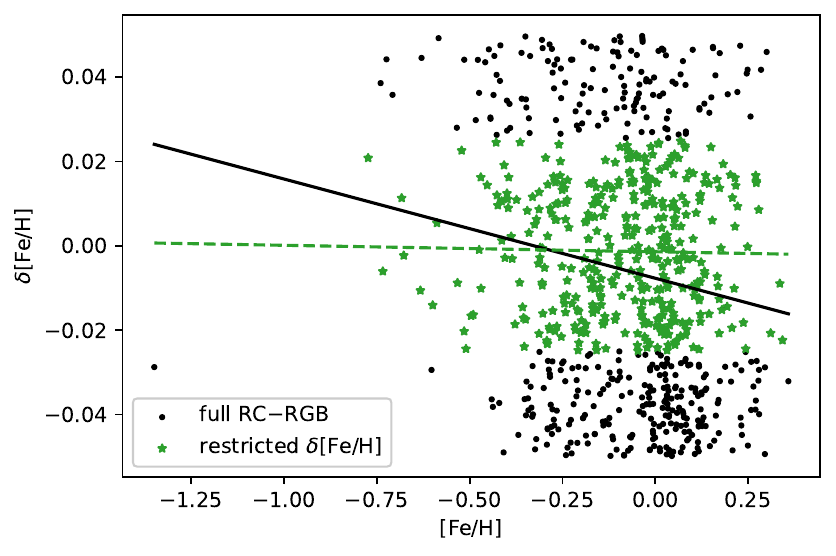}
    \includegraphics[width=0.49\linewidth]{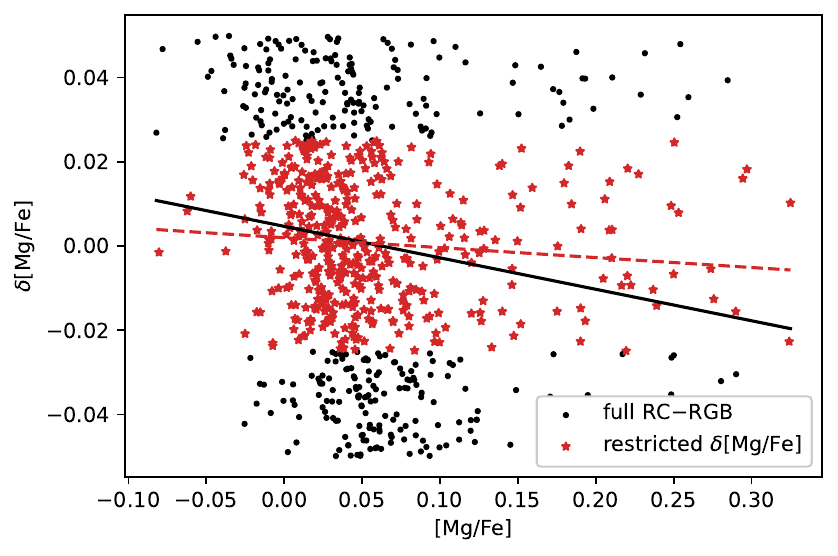}
    \caption{The correlation between RC stellar parameters and the stellar parameter difference, where each panel shows a different stellar parameter. The full RC$-$RGB matched set is shown in black whilst the restricted matched sets are shown in different colours as well as panels: restricted $\teff$ is shown in the top left (blue), restricted $\logg$ is shown in the top right (orange), restricted $\feh$ is shown in the bottom left (green), and restricted $\mgfe$ is shown in the bottom right (red). The solid line is a linear fit for each parameter and matched set, showing that the gradient in the restricted sets are reduced compared to the full set.}
    \label{fig:trend}
\end{figure*}

There is a correlation between stellar parameter and stellar parameter difference, as shown in Fig.~\ref{fig:param_hist}. We fit a linear line to the RC$-$RGB matched set to show the gradient of this trend in Fig.~\ref{fig:trend}. To study the impact of this correlation and distribution in stellar parameter difference on the median difference spectrum, we create a new matched set per stellar parameter, where the stellar parameter difference restriction is smaller by a factor of 2 compared to the RC$-$RGB matched set. The correlation between stellar parameter and stellar parameter difference of these new matched sets are shown in Fig.~\ref{fig:trend}. We fit a linear line to the new matched sets, showing that the gradient is reduced compared to the full RC$-$RGB matched set. The mean stellar parameter difference for the new matched sets are reported in Table~\ref{tab:res_delta_sp}, showing that our new matched sets have a reduced mean stellar parameter difference for the parameter that is controlled, whilst the mean difference between the other parameters do not change significantly. 

\begin{table*}
    \centering
    \caption{The mean difference in stellar parameters for the full RC$-$RGB matched set and the restricted matched sets. The mean difference for the restricted matched set is reduced compared to the full RC$-$RGB matched set for the corresponding stellar parameter, whilst the other mean stellar parameter differences of the non-restricted stellar parameters remains at a similar magnitude, limiting its impact on the median difference spectrum.}
    \begin{tabular}{c|cccccc}
\hline
Parameter & Units & RC$-$RGB & $\delta\teff$ & $\delta\logg$ & $\delta\feh$ & $\delta\mgfe$ \\
\hline 
\hline
$\teff$ & K & $7.17$ & $1.79$ & $6.56$ & $4.87$ & $7.25$ \\
$\logg\ /\ 10^{-3}$ & $\log(\cms)$ & $-7.5$ & $-8$ & $0.59$ & $-7.1$ & $-7.9$ \\
$\feh\ /\ 10^{-3}$ & dex & $5.8$ & $3.5$ & $5.8$ & $1.3$ & $5.5$ \\
$\mgfe\ /\ 10^{-4}$ & dex & $-3.5$ & $-1.3$ & $-0.71$ & $-3.3$ & $-6.5$ \\
\hline
    \end{tabular}
    \label{tab:res_delta_sp}
\end{table*}

\begin{figure*}
    \centering
    \includegraphics[width=\linewidth]{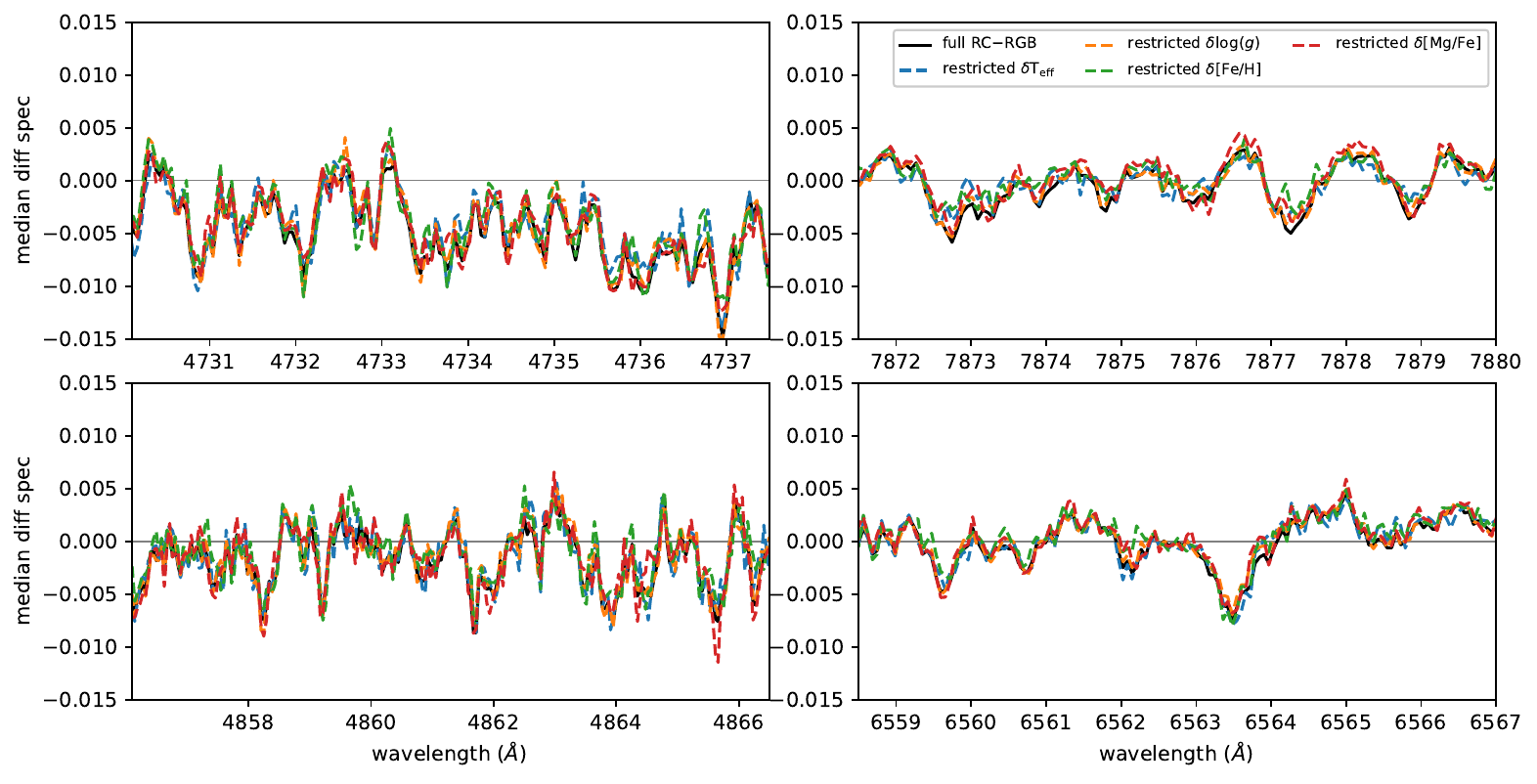}
    \caption{The median difference spectra: RC$-$RGB (solid black), restricted $\delta\teff$ (dashed blue), restricted $\delta\logg$ (dashed orange), restricted $\delta\feh$ (dashed green), and restricted $\delta\mgfe$ (dashed red). The difference between the full RC$-$RGB median difference spectrum and the restricted median difference spectrum is minimal for the wavelength regions shown, where we show the same wavelength regions as in Section~\ref{sec:results}. These are the same restricted matched sets as in Fig.~\ref{fig:trend}.}
    \label{fig:impact}
\end{figure*}

We show the median difference spectrum for each difference restricted set and the original RC$-$RGB matched set in Fig.~\ref{fig:impact}. Whilst the median difference spectrum is closer to zero at some wavelengths (e.g. the $\feh$ restricted matched set at 7875.9\,\AA), the impact of lowering the mean stellar parameter difference and correlation between stellar parameter and stellar parameter difference is minimal. 

\subsection{Misclassifications}
\begin{figure*}
    \centering
    \includegraphics[width=\linewidth]{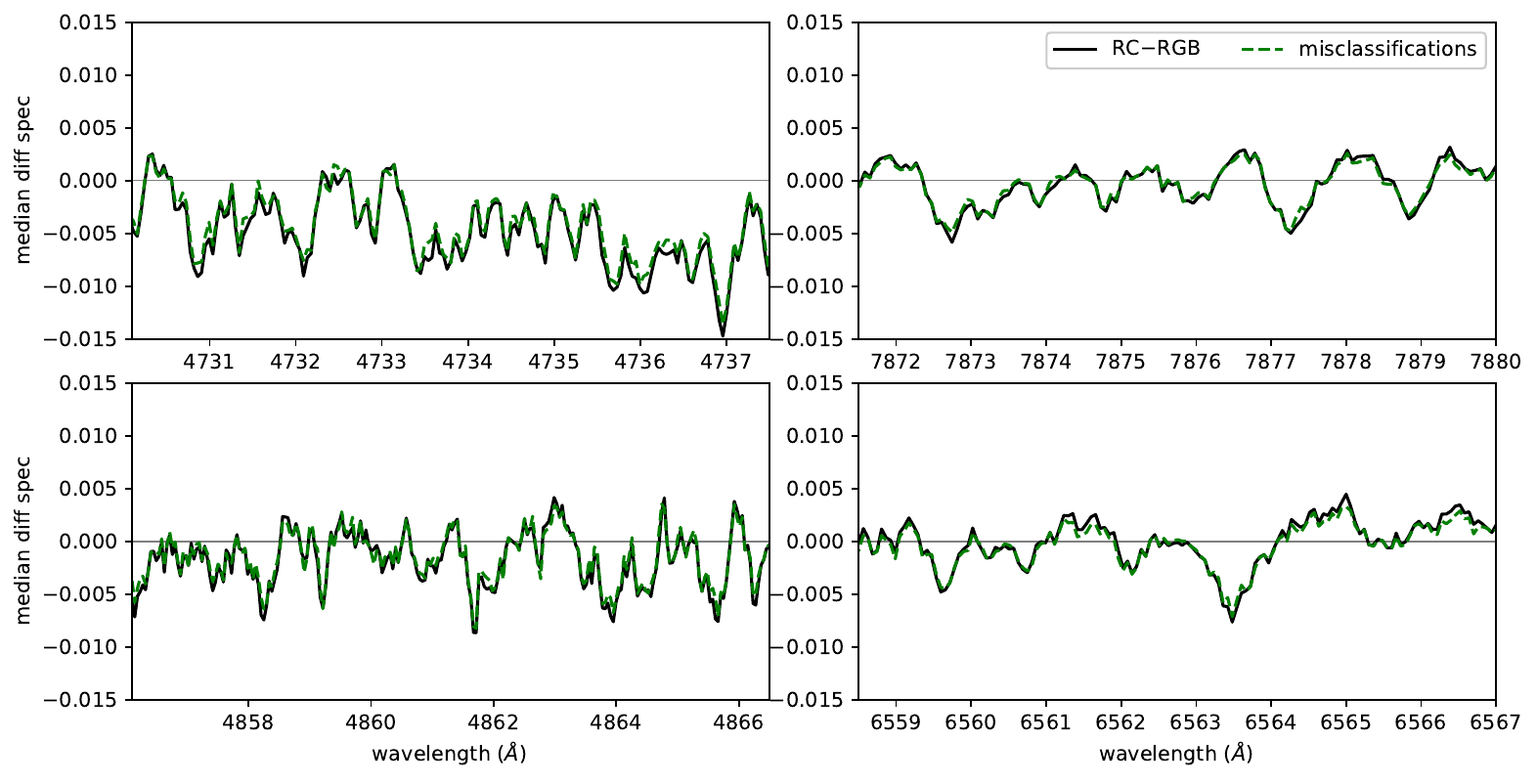}
    \caption{The median difference spectrum for the RC$-$RGB matched set (black) and a set of matched stars with 10\% misclassifications (green). The difference between the median different spectra is minimal, indicating that 10\% misclassified RGB stars do not affect the median difference spectrum.}
    \label{fig:inj}
\end{figure*}

Whilst we adopt asteroseismic classifications, they still could contain misclassifications. In order to study the effect of mislcassifications on our median difference spectrum, we create a new matched set with some misclassifications injected into the data. For this misclassified matched set, we take 10\% of our RC$-$RGB matched pairs and find a new pair, except this time with the same evolutionary state. We set our misclassification percentage to 10\% based on the random forest accuracy from Appendix~\ref{app:rf}. Fig.~\ref{fig:inj} shows the median difference spectrum for this misclassified matched set, which is similar to the RC$-$RGB matched set, indicating that misclassifications do not significantly affect our results. 

\section{Uncertainty of difference spectrum}
\label{app:stats}

Each spectrum of \galah has flux error at every wavelength. We are interested in quantifying the error in the median difference spectrum due to this flux error. In the following, we derive Eq.~\ref{eq:err}.

Let $f_{i}$ be the spectrum for the $i$th star, then for a total of $n$ stars, $i \in {1, \ldots ,n}$. We stack the spectra by taking the median: $f_m = {\rm med}(f_{1}, \ldots, f_{n})$. Assuming the spectra are independent, the errors can then be combined in terms of their $\snr$ by summation in quadrature:
\begin{equation}
    \label{eq:combine_snr}
    [{\snr}](f_m) = \sqrt{\sum_{i=1}^n [{\snr}]^2(f_i)},
\end{equation}

Note that the flux error for spectrum $f_{i}$ is the inverse of the $\snr$,
\begin{equation}
    \label{eq:snr_to_sigma}
    \sigma_i = \frac{1}{[\snr](f_i)}.
\end{equation}

Then the error of the median spectrum is given by the combination of Eq.~\ref{eq:combine_snr} and \ref{eq:snr_to_sigma}:
\begin{equation}
    \label{eq:stack_error}
    \sigma_m = \left( \sum_{i=1}^n \frac{1}{\sigma^2_i} \right)^{-\frac{1}{2}}.
\end{equation}

In this work, we take the median of the difference spectra, $\delta f_i = f_{{\rm RC},i} - f_{{\rm RGB},i}$. The error in the difference spectrum is given by the additivity of variance:
\begin{equation}
    \label{eq:ind_var_sum}
    \sigma^2_i = \sigma^2_{{\rm RC},i} + \sigma^2_{{\rm RGB},i}.
\end{equation}
therefore, by combining Eq.~\ref{eq:stack_error} and Eq.~\ref{eq:ind_var_sum} we reach Eq.~\ref{eq:err}.

\section{Random Forest}
\label{app:rf}
\edit{In an effort to expand the number of stars with asteroseismic classifications, machine learning classification algorithms are often employed to transfer the evolutionary state label from asteroseismic datasets to spectroscopic datasets \citep{hawkins2018, ting2018, casey2019, lu2022}. Random forests are often employed for this purpose due to their speed of training and high accuracy. In this appendix section, we apply random forests to our crossmatched catalogue training on different combinations of both stellar parameters and spectra, to identify which features contribute to the evolutionary state classification. }

\edit{For these tests we, utilise our full crossmatched catalogue which contains a total of \total stars, where we adopt a 80--20 train-test cross validation split for evaluation of our random forest performance. We use \texttt{RandomForestClassifier} from the \textsc{scikit-learn} library \citep{scikit-learn}, and optimise hyperparameters using \texttt{GridSearchCV} over hyperparameter ranges:
\begin{itemize}
    \item Criterion utilized: $\{$ \texttt{gini}, \texttt{entropy} $\}$;
    \item Max depth of a single tree: $\{ 5, 6, 7, 8, 9, 10, 11, 12, 13, 14, 15 \}$;
    \item Number of estimators of the forest: $\{ 10, 100, 500, 1000, 5000, 10000, 20000 \}$;
    \item Maximum samples utilized in training a single tree: $\{ 0.3, 0.5, 0.7, 0.9 \}$.
\end{itemize}}

\edit{We find that when training on stellar parameters it is difficult to improve the accuracy beyond 91\%, which is achieved through training on $\teff$, $\logg$, $\feh$; likely due to the low accuracy of additional features such as $\vmic$, $\cn$ on a star-by-star basis.}
When learning solely on spectra, the random forest finds evolutionary state information in every wavelength of spectra\edit{, indicating that there is evolutionary state information in spectra beyond just the $\ctwo$ and CN molecular lines. }

\subsection{Stellar parameters}
We train random forests on different combinations of stellar parameters to classify the evolutionary state. \edit{The combination of stellar parameters trained on and the accuracy achieved are listed in Table~\ref{tab:rf_sp}.} The stellar parameters $\teff$, $\logg$, and $\feh$ can provide a classification accuracy of 91\% for our crossmatched catalogue. Out of these three stellar parameters, $\logg$ contains the most information, with classification accuracy reaching 81\% when trained on $\logg$ alone. \edit{It may seem surprising that the accuracy when training on $\logg$ alone is relatively high, but this is expected, as RC stars are concentrated at $\logg = 2.5$.} The inclusion of $\teff$ and $\feh$ to classification using $\logg$ both provide an increase of $\sim 5\%$ in accuracy. Classifying on $\vmic$ and the stellar parameters provides an increase of 1\% to the accuracy, whilst classifying on $\cn$ or $\vsini$ offers no improvement to classification accuracy. This result is expected as, although we observe $\vmic$, $\cn$, and $\vsini$ differences for the full population of RC and RGB stars, the accuracy and precision for individual stars are not enough to classify evolutionary state accurately on a star-by-star basis. \edit{In general, the accuracy improves as more features are used to train the random forest. However, this improvement reaches a limit around 91\%, where it becomes difficult to improve the accuracy further. }

\begin{table}
    \centering
    \caption{Different combinations of features used to train the random forest and the accuracy achieved by the trained random forest.}
    \begin{tabular}{c|c|c|c|c|c|c|c}
\hline
$\teff$ & $\logg$ & $\feh$ & $\vsini$ & $\cn$ & $\vmic$ & Accuracy \\
\hline 
\hline
\checkmark & \xmark & \xmark & \xmark & \xmark & \xmark & 70\% \\
\xmark & \checkmark & \xmark & \xmark & \xmark & \xmark & 81\% \\
\checkmark & \checkmark & \xmark & \xmark & \xmark & \xmark & 85\% \\
\checkmark & \checkmark & \checkmark & \xmark & \xmark & \xmark & 91\% \\
\checkmark & \checkmark & \checkmark & \checkmark & \xmark & \xmark & 91\% \\
\checkmark & \checkmark & \checkmark & \xmark & \checkmark & \xmark & 91\% \\
\checkmark & \checkmark & \checkmark & \xmark & \xmark & \checkmark & 92\% \\
\checkmark & \checkmark & \checkmark & \xmark & \checkmark & \checkmark & 92\% \\
\hline
    \end{tabular}
    \label{tab:rf_sp}
\end{table}

We find $78\%$--$94\%$ of incorrectly classified stars are common between random forests trained on different combinations of features. That is, if a star is incorrectly classified through a particular combination of features, it is also likely to be incorrectly classified through a different combination of features. Therefore, these accuracy percentages are highly dependent on the dataset. It is possible that these stars have an incorrect evolutionary state label from our asteroseismic catalogue, however, we do not see a correlation between incorrect classification by the random forest and the evolutionary state \edit{classification} or $\dnu$ \edit{vetting}. 

\subsection{Stellar spectra}
There is evolutionary state information in every wavelength in spectra. We trained two hyperparameter optimised random forests on the Swan bands. One random forest used $\sim350$ wavelengths with $\ctwo$ molecular features, whilst the other random forest used $\sim350$ wavelengths without $\ctwo$ molecular features. Both random forests achieved an accuracy of $\sim 84\%$, indicating that there is evolutionary state information outside of the $\ctwo$ molecular features. In Section~\ref{sec:results} \edit{and~\ref{sec:discussion}}, we find that RC stars have broader \edit{\ha and \hb} than RGB stars, linked to the $\vmic$ parameter. Line-width information is present at every wavelength, therefore, it is possible that the random forest is classifying based on differences in line-width present in every line in the spectrum. \edit{Despite these results, identifying the exact features that the random forest uses to make its predictions requires further careful study due to the interpretability issues associated with large random forest models~\citep{haddouchi2019survey}.
}

\bsp	
\label{lastpage}
\end{document}